\documentclass[12pt]{article}

\usepackage{bm}
\usepackage{float}
\usepackage{relsize}
\usepackage{array}
\usepackage{fullpage}
\usepackage{amsfonts}
\usepackage{amsmath}
\usepackage{setspace}
\usepackage{easybmat}
\usepackage{slashed}
\usepackage{amsmath}
\usepackage{amssymb}
\usepackage{graphicx}
\usepackage{verbatim}
\usepackage{caption}
\usepackage{subcaption}
\usepackage{hyperref}

\allowdisplaybreaks
\def\timesbox{\hbox{$\scriptscriptstyle\times$}}
\def\ant{ {{\lower 1ex  \timesbox} \atop {\raise 1.5ex  \timesbox}}}
\def\f{\frac}
\def\pa{\partial}
\def\non{\nonumber\\}
\def\b{\beta}

\def\d{\delta}
\def\e{\epsilon}
\def\g{\gamma}

\def\l{\lambda}
\def\m{\mu}
\def\n{\nu}

\def\p{\pi}

\def\r{\rho}
\def\s{\sigma}
\def\t{\tau}
\def\th{\theta}
\def\D{\Delta}

\def\G{\Gamma}

\def\O{\Omega}

\newcommand{\mathsym}[1]{{}}
\newcommand{\unicode}[1]{{}}

\newcommand{\be}{\begin{equation}}
\newcommand{\ee}{\end{equation}}
\newcommand{\beqa}{\begin{eqnarray}}
\newcommand{\eeqa}{\end{eqnarray}}
\newcommand{\bsp}{\begin{split}}
\newcommand{\esp}{\end{split}}
\newcommand{\bgth}{\begin{gather}}
\newcommand{\egth}{\end{gather}}
\newcommand{\ra}{\rangle}
\newcommand{\la}{\langle}

\newcommand{\tr}{\hbox{tr}}

\begin{document}

\title{\textbf{On partition functions and phases of scalars in AdS}
\\
\author{Astha Kakkar$^{a}$\footnote{asthakakkar8@gmail.com} and 
Swarnendu Sarkar$^{a,b}$\footnote{swarnen@gmail.com}\\
$^a$\small{{\em Department of Physics and Astrophysics,
University of Delhi,}} \\ 
\small{{\em Delhi 110007, India}}\\
$^b$\small{{\em Department of Physics, Vidyasagar University,}}\\
\small{{\em Midnapore 721102, India}}\\
}}

\date{}
\maketitle

\abstract{We study the phases of scalar field theories in thermal $AdS_{d+1}$ spaces for $d=1,2,3$. The analysis is done for theories with global $O(N)$ symmetry for finite as well as large $N$. The symmetry-preserving and symmetry-breaking phases are identified as a function of the mass-squared of the scalar field and temperature. On the way we also describe a method for computing one-loop partition function for scalar field on thermal $AdS_{d+1}$ for arbitrary $d$ that reproduces results known in the literature. The derivation is based on the method of images and uses the eigenfunctions of the Laplacian on Euclidean $AdS$.}

\newpage

\tableofcontents

\baselineskip=18pt

\section{Introduction}\label{intro}

The primary ingredient for studying the phases of a quantum field theory is the effective action. Though obtaining an exact form is beyond the scope of the existing techniques, approximate expressions using perturbative methods are available. To the leading order this  involves computation of one-loop determinants. In this paper we employ these results for scalar field theories in thermal $AdS$. Specifically we shall be interested in studying the phases of scalar theories as a function of the mass-squared $(m^2)$ of the field and temperature $(T=1/\b)$ in various dimensions.

One-loop partition functions for quantum fields on maximally symmetric spaces have been the topic of interest for quite some time \cite{Burgess:1984ti}-\cite{Das:2006wg}. Since the initial computations, various new techniques have been invented. The renewed momentum in the field was primarily motivated from the AdS/CFT correspondence. Thermal $AdS_{d+1}$ is the quotient $\mathbb{H}^{d+1}/\mathbb{Z}$. Using this fact, expressions for the partition functions were obtained for $AdS_3$ after computing the heat kernels using the method of images in \cite{Giombi:2008vd}. By noting that the one-loop determinant is a meromorphic function of $\D=d/2+\sqrt{(d/2)^2+m^2}$ the authors of \cite{Denef:2009kn} have obtained the corresponding expressions utilizing the normal modes. Group theoretic methods were used in \cite{David:2009xg}-\cite{Gopakumar:2011qs} to obtain the heat kernels for $d\ge 3$. Connection between the heat kernel methods and that using normal modes is recently explored in \cite{Martin:2019flv}. In this paper we give an alternate derivation of these results for scalars using the eigenfunctions of the Laplacian operator in $AdS$ for Euclidean $AdS$. We show that our computation essentially involves the method of images applied to the Green's function and generalizes to $AdS$ spaces with arbitrary dimensions. A similar method using global coordinates was used in \cite{Kraus:2020nga}. The computation here being done using Poincar\'e coordinates differs from this in details. 

The subsequent sections of the paper are devoted to the study of phases of scalar theories on thermal $AdS_{d+1}$. We identify the regions in the $\b$-$m^2$ parameter space that correspond to the symmetry preserving and symmetry breaking phases for $d=1,2,3$. 
It is expected that the ultraviolet properties of the theory remain unaffected by the finite curvature of $AdS$ space. As a consequence the zero temperature contribution to the leading effective action is divergent in the ultraviolet as in flat space for $d \ge 1$. We regularize this using dimensional regularization and obtain a finite answer by using the standard procedure of adding counterterms. 
It is further essential to regularize the volume of thermal $AdS_{d+1}$ in order to proceed with the analysis. 
We use two regularization schemes namely cutoff \cite{Sen:2008vm}-\cite{Cotler:2018zff} and dimensional regularization \cite{Graham:1999jg}-\cite {Diaz:2007an} to obtain the regularized volume ${\cal V}_{d+1}$. The sign of ${\cal V}_{d+1}$ plays a crucial role in the qualitative nature of the phase diagrams. For $AdS_3$ the regularized volume ${\cal V}_{3}$ from the two schemes differ in magnitude and in sign. The zero temperature contribution to the effective potential is finite and unambiguous. For example, both the dimensional regularization and the heat kernel methods give the same result. We thus study the phases of the $O(N)$ vector model corresponding to both the regularization schemes, each giving a different qualitative behavior. The choice of the phase diagram is presumably determined by fixing the ground state energy of the boundary CFT.  Along the same lines, we find that the phase diagrams for the $O(N)$ vector model, with finite $N$ on $AdS_2$ and $AdS_3$ with negative ${\cal V}_{d+1}$ are similar and differ from that on $AdS_4$ and $AdS_3$ having ${\cal V}_{d+1}$ positive. An interesting feature in $d=1,2$, when ${\cal V}_{d+1}$ is negative is that, the symmetry breaking phase persists at high temperatures unlike that in flat space.   

For the large $N$, $O(N)$ vector model, one of the constraints that is important in the determination of the phases is the existence of solution corresponding to the saddle point equation. In our analysis we are able to isolate a region where there is no solution corresponding to an extremum of the potential for some of the cases. We suspect that this is an artifact of the perturbative analysis. In $AdS$ space unlike that of flat space, there is a region in the parameter space where both the phases coexist. This is due to the fact that $AdS$ space allows the scalars to have negative mass-squared up to $-(d/2)^2$, which is the Breitenlohner-Freedman (BF) bound. The preferred phase of the system is determined by the one with a lower value of the potential. Such a phase was identified at zero temperature in \cite{Carmi:2018qzm}. We see here that the phase extends into lower values of $\b$ (higher temperatures). We further find a limiting temperature $\b_l$ below which this phase ceases to exist. Another departure in the phase structure here from that in flat space is the occurrence of the symmetry breaking phase in two dimensions. This was noted in \cite{Inami:1985dj} for the large $N$, $O(N)$ vector model. It was argued that this is due to the curvature of $AdS$ and should persist for finite $N$. The analysis here confirms this for finite temperatures.    

The paper is organized as follows. In section \ref{eff} we study the computation of one-loop determinant for scalar field in $AdS_{d+1}$.  We first review the computation for Euclidean $AdS_{d+1}$ in section \ref{euclid}.  In section \ref{thermal} we give a derivation of the partition function for thermal $AdS_{d+1}$. The phases of scalars are studied in section \ref{phases}. We study the phases for a theory with a single scalar field in section \ref{single} and that of the $O(N)$ vector model in section \ref{onmodel} for various dimensions. Large $N$, $O(N)$ vector model is studied in section \ref{largen}. We summarize in section \ref{summary}. In appendix \ref{var} we present details that demonstrate equality of the various integral representations of the trace given in section \ref{euclid}. In appendix \ref{del} we deduce delta-function identities used in section \ref{thermal}.   

\section{One-loop effective action at zero and finite temperature }\label{eff}

We begin with some basics to setup the notations.
The action for a real scalar field on an Euclidean ${d+1}$ dimensional space is 

\begin{equation}
S=-\int d^{d+1}x \sqrt{g}\left[\frac{1}{2}(\partial^{}_{\mu}\phi)^{2}+V(\phi)\right].
\end{equation}

Expanding about the constant classical value of $\phi$, which extremizes the potential, as $\phi=\phi^{}_{cl}+\eta$ and integrating over $\eta$, one gets the one-loop partition function,

\begin{eqnarray}
Z&=&\det[-\square^{}_{E}+V''(\phi^{}_{cl})]^{-1/2} \exp(-{\cal V}^{}_{d+1} V(\phi^{}_{cl}))\non 
&=&\exp\left(-\frac{1}{2}\mbox{tr} \log[-\square^{}_{E}+V''(\phi^{}_{cl})]\right)\exp\left(-{\cal V}_{d+1} V(\phi^{}_{cl})\right)
\end{eqnarray}

where, 

\begin{equation}
\square^{}_{E}=\frac{\partial^{}_{\mu}[\sqrt{g}g^{\mu\nu} \partial^{}_{\nu}]}{\sqrt{g}}
\end{equation}

and ${\cal V}^{}_{d+1}$ is the volume of $d+1$ dimensional Euclidean space.
The effective potential is thus,
\begin{equation}
\label{c1e1}
V^{}_{eff}(\phi^{}_{cl})=-\frac{1}{{\cal V}^{}_{d+1}}\log Z=\frac{1}{2{\cal V}^{}_{d+1}}\mbox{tr} \log[-\square^{}_{E}+V''(\phi^{}_{cl})]+V(\phi^{}_{cl}).
\end{equation}
Extremizing the effective potential
\begin{equation}
\label{c1e2}
\frac{dV^{}_{eff}}{d\phi^{}_{cl}}=V'(\phi^{}_{cl})+\frac{1}{2{\cal V}^{}_{d+1}}\mbox{tr}\left[ \frac{1}{-\square^{}_{E}+V''(\phi^{}_{cl})}\right] V'''(\phi^{}_{cl}).
\end{equation} 

We shall be computing the trace in the above equation which can also be written as $\pa \log Z^{(1)}/\pa M^2$ with 
\beqa
\log Z^{(1)}=-\frac{1}{2}\mbox{tr} \log[-\square^{}_{E}+M^2]
\eeqa

\subsection{Euclidean $AdS_{d+1}$}\label{euclid}

In this section we review the computation of the trace in (\ref{c1e2}) for Euclidean $AdS_{d+1}$. The metric of $AdS_{d+1}$ in Poincar\'e coordinates is:

\beqa
ds^2=\f{L^2}{y^2}\left(dy^2+\eta_{\mu\nu}dx^{\mu}dx^{\nu}\right).
\eeqa

In these coordinates we first solve the following eigenvalue equation,

\beqa\label{eeqn}
-L^2 \Box_E \Psi_{\l,\vec{k}}(\vec{x},y)=\left[\l^2+\left(\frac{d}{2}\right)^2\right] \Psi_{\vec{k},\l}(\vec{x},y)
\eeqa

where,

\beqa\label{box}
-L^2 \Box_E=-y^2\left[y^{d-1}\pa_y\left(y^{1-d}\pa_y\right)+\pa_{\vec{x}}^2\right].
\eeqa

Writing $\Psi_{\vec{k},\l}(\vec{x},y)=\phi_{\l}(y)e^{\pm i \vec{k}.\vec{x}}$, and defining 
$\phi_{\l}(y)=(ky)^{d/2}\tilde{\phi}_{\l}(ky)$, the above equation leads to the following differential equation

\beqa
\tilde{y}^2\tilde{\phi}_{\l}^{\prime\prime}(\tilde{y})+\tilde{y}\tilde{\phi}_{\l}^{\prime}(\tilde{y})+[\l^2-\tilde{y}^2]\tilde{\phi}_{\l}(\tilde{y})=0 
\eeqa

where $\tilde{y}=ky$. The solution to this equation gives the eigenfunctions of (\ref{eeqn}). They are given by the modified Bessel function of the second kind,  
$\tilde{\phi}_{\l}(ky)=K_{i\l}(ky)$ that is regular as $y\rightarrow \infty$ and thus ${\phi}_{\l}(ky)=(ky)^{d/2}K_{i\l}(ky)$. A scalar field $\phi$ of mass $m$ is dual to the operator in CFT whose conformal dimension $\D$ satisfies $\D(\D-d)=(mL)^2$. Out of the two roots, $\D_{\pm}=d/2\pm\sqrt{(d/2)^2+(mL)^2}$, of this equation, this choice of the solution, corresponds to choosing $\D=\D_{+}>d/2$.

The trace in (\ref{c1e2}) can be computed using the spectral form as follows

\beqa\label{trace1}
&&\frac{1}{L^2}\tr\left[\frac{1}{- \Box_E+V^{''}(\phi_{cl})}\right]\non
&=&\f{1}{L^{2}}\int d^{d+1}x \sqrt{g}\int d\l ~\m(\l)\int \frac{d^d k}{(2\pi)^d}\f{1}{(L^{d+1}k^d)}\langle\l,k|\left[\frac{1}{- \Box_E+V^{''}(\phi_{cl})}\right]|y,\vec{x}\rangle\langle y,\vec{x}|\l,k\rangle\non
&=&\f{1}{L^{d+1}}\int d^{d+1}x \sqrt{g}\int  \f{d\l~\m(\l)}{\l^2 +\n^2}\int \frac{d^d k}{(2\pi)^d} y^d K^2_{i\l}(ky)
\eeqa

where $\n=\sqrt{\left(\f{d}{2}\right)^2 + L^2 V^{''}(\phi_{cl})}$ and 

\beqa
\m(\l)=\f{2 \l}{\p^2}\sinh(\p\l)
\eeqa

We have used the following normalizations,

\beqa
\int d^{d+1}x \sqrt{g} |x\rangle\langle x|&=&1~~~~~\mbox{;}~~~~~|x\rangle=|\vec{x}\rangle \otimes |y\rangle\non
%\langle \vec{k}|\vec{k}^{'}\rangle &=& (2\p)^d\d^d(\vec{k}-\vec{k}^{'})\non
\langle y,\vec{x}|\l,\vec{k}\rangle&=&e^{i\vec{k}.\vec{x}}\left(ky\right)^{d/2}K_{i\l}(ky)\label{1form}
%\langle z|\l\rangle&=& K_{i\l}(z)
\eeqa
  
The measures $\m(\l)$ and $1/k^{d}$ are obtained from 

\beqa
\la \l,\vec{k}|\l^{\prime},\vec{k}^{'}\ra &=& \int d^{d+1}x \sqrt{g}~ e^{-i(\vec{k}-\vec{k}^{\prime}).\vec{x}}\left(ky\right)^{d/2}\left(k^{\prime}y\right)^{d/2}K_{i\l}(ky) 
K_{i\l^{\prime}}(k^{'}y)\non
&=&  L^{d+1}k^d(2\p)^d\d^d(\vec{k}-\vec{k}^{\prime})\int \f{dy}{y}K_{i\l}(ky) 
K_{i\l^{\prime}}(k^{\prime}y)\non
&=& L^{d+1}k^d(2\p)^d\d^d(\vec{k}-\vec{k}^{\prime})\f{\d(\l-\l^{\prime})}{\m(\l)}
\eeqa  
so that 

\beqa
\int  \frac{d^d k}{(2\pi)^d}\f{1}{(L^{d+1}k^d)} \int d\l ~\m(\l) \la \l,\vec{k}|\l^{'},\vec{k}^{'}\ra =1
\eeqa

The trace can be computed by performing the integrals in various orders. One can first integrate over $k$ in (\ref{trace1}) and further perform the integral over $\l$ by closing the contour in the upper half of the complex $\l$ plane. This has been done in Appendix \ref{var}.  

Other representations of the trace can obtained by using the following identities \cite{Grad}(see also \cite{Miyagawa:2015sql}):

\beqa\label{ikrel}
K_{i\l}(ky)=\f{\p}{2}\f{I_{-i\l}(ky)-I_{i\l}(ky)}{i\sinh(\p\l)}
\eeqa

and for $y< y^{\prime}$,

\beqa
I_{-i\l}(ky)K_{i\l}(ky^{\prime})&=& \int_0^{\infty} \f{ds}{2s}e^{-k^2 s} e^{-\f{y^2+y^{\prime 2}}{4s}}I_{-i\l}\left(\f{yy^{\prime}}{2s}\right)\non
&=& \int_0^{\infty} \f{ds}{2s}e^{-k^2 s} e^{-\f{y^2+y^{\prime 2}}{4s}} \frac{1}{2\p i}\int_{\infty-i\p}^{\infty+i\p} dt \exp\left(\frac{yy^{\prime}}{2s}\cosh (t)+i\l t\right)
\eeqa

one can write

\beqa
\frac{1}{L^2}\tr\left[\frac{1}{- \Box_E+V^{''}(\phi_{cl})}\right]&=& \f{1}{L^{d+1}}\int d^{d+1}x \sqrt{g}  y^d  \int \frac{d^d k}{(2\pi)^d}\f{1}{(2\pi i)}\int d \l \frac{2\l}{\l^2+\n^2}\\
&\times& \left[\lim_{y^{\prime}\rightarrow y}\int\f{ds}{2s}e^{-k^2 s} e^{-\f{y^2+y^{\prime 2}}{4s}} \frac{1}{2\p i}\int_{\infty-i\p}^{\infty+i\p} dt \exp\left(\frac{yy^{\prime}}{2s}\cosh (t)+i\l t\right)\right]\nonumber
\eeqa

which allows one to perform the contour integral over $\l$ in the upper half plane. This leads to the trace being written as

\beqa
&&\f{1}{L^{d+1}}\int d^{d+1}x \sqrt{g} ~ y^d  \int \frac{d^d k}{(2\pi)^d}\left[\lim_{y^{\prime}\rightarrow y}\int\f{ds}{2s}e^{-k^2 s} e^{-\f{y^{2}+y^{\prime 2}}{4s}}\frac{1}{2\p i}\int_{\infty-i\p}^{\infty+i\p} dt \exp\left(\frac{yy^{\prime}}{2s}\cosh (t)-\n t\right)\right]\non\label{order3}\\
&=&\f{1}{L^{d+1}}\int d^{d+1}x \sqrt{g} ~ y^d  \int \frac{d^d k}{(2\pi)^d}\left[\lim_{y^{\prime}\rightarrow y}\int \f{ds}{2s}e^{-k^2 s} e^{-\f{y^{2}+y^{\prime 2}}{4s}}I_{\n}\left(\f{yy^{\prime}}{2s}\right)\right]\label{order2}\\
&=& \f{1}{L^{d+1}}\int d^{d+1}x \sqrt{g} ~ y^d  \int \frac{d^d k}{(2\pi)^d} \left[\lim_{y^{\prime}\rightarrow y}I_{\n}(ky)K_{\n}(ky^{\prime})\right]\label{ikform} \\
&=&\f{{\cal V}_{d+1}}{L^{d+1}}\f{\G\left(d/2+\n\right)\G(1/2-d/2)}{\G\left(1-d/2+\n\right) (4\p)^{(d+1)/2}}\label{trzt}
\eeqa

The above expression was derived in \cite{Burgess:1984ti,Inami:1985wu} and the corresponding effective actions have been studied in various works (see for example \cite{Burgess:1984ti}-\cite{Kamela:1998mb},\cite{Gubser:2002zh},\cite{Das:2006wg},\cite{Carmi:2018qzm}). Equation (\ref{trzt}) gives the analytically continued result valid for $d \ge 1$ which is otherwise (ultraviolet) divergent for $d\ge 1$. Alternate approach such as the zeta-function regularization was used in \cite{Camporesi:1990wm}-\cite{Caldarelli:1998wk}.  

The integrals in (\ref{order3}) and (\ref{order2}) can also be performed separately, the results for both of which can be brought to the form (\ref{trzt}). This is shown in Appendix \ref{var}. This exercise serves as a check for the various integral representations of the trace. The final form given by (\ref{ikform}) will be used in computation of the trace on thermal $AdS$.

\subsection{Thermal $AdS$}\label{thermal}

In this section we compute the one-loop partition function for scalar field theory in the thermal $AdS$ background. The expression will be used to study the phases of scalar field theories in section \ref{phases}. As mentioned before the result that we obtain is well known in the literature. Here we give an alternate derivation of the finite temperature contribution using the spectral form of the trace. The computation is based on the method of images and uses the eigenfunctions already derived for the Euclidean space\footnote{A similar method was used to compute entanglement entropy in $AdS$ in \cite{Miyagawa:2015sql},\cite{Sugishita:2016iel}.}. We begin by illustrating the method for thermal $AdS_3$ and then generalize this for $AdS_{d+1}$ for $d$ even. We shall further see that the derivation generalizes to even dimensional $AdS$ spaces.

\subsubsection{Thermal $AdS_3$}

Thermal $AdS_3$ is the quotient space $\mathbb{H}^3/\mathbb{Z}$ having the metric 
\beqa
ds^2=\f{L^2}{y^2}\left(dy^2+dzd\bar{z}\right)
\eeqa

with the action of $\g^{n}\in \mathbb{Z}$ on the coordinates as,

\beqa
\g^n(y,z)=(e^{-n\beta}y, e^{2\pi in\tau}z) ~~~~\mbox{where}~~~~\tau=\frac{1}{2\pi}(\theta+i\beta)
\eeqa

In terms of real coordinates $z=x_1+ix_2$, the action of the group element on the real coordinates $\vec{x}=(x_1,x_2)$ can be written as, 

\beqa\label{realt}
\g^n \vec{x}=\left(e^{-n\beta}(x_1 \cos n\theta-x_2\sin n\theta),e^{-n\beta}(x_1 \sin n\theta+x_2\cos n\theta)\right)
\eeqa

Using the eigenfunctions $\Psi_{\l,\vec{k}}(\vec{x},y)$ that solve (\ref{eeqn}), we write down the following function that is invariant under the above action on the coordinates,

\beqa\label{norma}
\Phi_{\vec{k},\l}(\vec{x},y)=\frac{1}{{\cal N}}\sum_{n=-\infty}^{\infty} (ke^{-n\beta}y)K_{i\l}(k e^{-n\beta}y)e^{-i\vec{k}.(\g^n \vec{x})}
\eeqa  

It is easier to perform the integrals over the full $\mathbb{H}^3$ instead of the fundamental domain of $\mathbb{H}^3/\mathbb{Z}$. This however results in the sum being divergent. The normalization ${\cal N}$ introduced in (\ref{norma}) regularizes the sum over $n$ as explained further below.

Similar to that of the zero temperature, and setting $L=1$, we normalize  $\Phi_{\vec{k},\l}(\vec{x},y)$ as follows,

\beqa\label{normt}
&&\int d^{3}x\sqrt{g}~\Phi_{\vec{k},\l}(\vec{x},y)\Phi^{*}_{\vec{k}^{\prime},\l^{\prime}}(\vec{x},y)\\\nonumber
&=&\frac{1}{{\cal N}^2}\sum_{n,n^{\prime}} \int d^{3}x\sqrt{g}~(ke^{-n\beta}y)(k^{\prime}e^{-n^{\prime}\beta}y)K_{i\l}(k e^{-n\beta}y)K_{i\l^{\prime}}(k^{\prime} e^{-n^{\prime}\beta}y)e^{-i\vec{k}.(\g^n \vec{x})}e^{i\vec{k}^{\prime}.(\g^{n^{\prime}} \vec{x})}\\\nonumber
&=&\frac{1}{{\cal N}^2}\sum_{n,n^{\prime}} \int \frac{dy}{y}~(ke^{-n\beta})(k^{\prime}e^{-n^{\prime}\beta})K_{i\l}(k e^{-n\beta}y)K_{i\l^{\prime}}(k^{\prime} e^{-n^{\prime}\beta}y)(2\pi)^2\d^2(\g^n \vec{k}-\g^{n^{\prime}} \vec{k}^{\prime})\\\nonumber
&=&\frac{1}{{\cal N}^2}\sum_{n,n^{\prime}} \int \frac{dy}{y}~(ke^{-(n-n^{\prime})\beta})^2K_{i\l}(k e^{-n\beta}y)K_{i\l^{\prime}}(k e^{-n\beta}y)(2\pi)^2\d^2(\g^{(n-n^{\prime})} \vec{k}-\vec{k}^{\prime})\\\nonumber
&=&\frac{1}{{\cal N}}\sum_{n} (ke^{-n\b})^2(2\pi)^2\d^2(\g^{n} \vec{k}-\vec{k}^{\prime})\d(\l-\l^{\prime})/\m(\l).
\eeqa

In the third line of equation (\ref{normt}) we have integrated over $\vec{x}$. The action of $\g^{n}$ on $\vec{k}$ is defined as,

\[
\g^n \vec{k}=\left(e^{-n\beta}(k_1 \cos n\theta+k_2\sin n\theta),e^{-n\beta}(k_2 \cos n\theta-k_1\sin n\theta)\right)
\]

so that $|\g^n \vec{k}|=e^{-n\b}k$. In the second-last line we have used the identity (see Appendix \ref{del})

\beqa\label{di1}
\d^2(\g^n \vec{k}-\g^{n^{\prime}} \vec{k}^{\prime})=e^{2n^{\prime}\b}\d^2(\g^{(n-n^{\prime})} \vec{k}-\vec{k}^{\prime})
\eeqa

In the last line we have used the fact that each value of $n^{\prime}$ gives the same series in $n$.  The sum over $n^{\prime}$ being divergent cancels a factor of ${\cal N}$ in the denominator. The divergence appears due to the infinite number of elements of the group $\mathbb{Z}$ (as in limit ${\cal N} \rightarrow \infty$ of $\mathbb{Z}_{\cal N}$).  
We shall see that we reproduce a finite expression for one-loop partition function at non-zero temperature.

The final line of (\ref{normt}) implies that,

\beqa
\frac{1}{{\cal N}}\sum_{n} \int  \frac{d^2 k}{(2\pi)^2}\f{1}{k^2} \int d\l ~\m(\l)(ke^{-n\b})^2(2\pi)^2\d^2(\g^{n} \vec{k}-\vec{k}^{\prime})\d(\l-\l^{\prime})/\m(\l)=1
\eeqa

We now compute the one-loop partition function. As in the zero temperature case, this is given by the first line of (\ref{trace1}) with the difference that, $\langle y,\vec{x}|\l,k\rangle=\Phi_{\vec{k},\l}(\vec{x},y)$. Noting that

\[\pa^2_{\vec{x}} e^{-i\vec{k}.(\g^n \vec{x})}=-|e^{-n\b}k|^2e^{-i\vec{k}.(\g^n \vec{x})}\]

we see that $\Phi_{\vec{k},\l}(\vec{x},y)$ have the same eigenvalues $\l$ as $\Psi_{\vec{k},\l}(\vec{x},y)$. We thus have

\beqa
&&\tr\left[\frac{1}{- \Box_E+V^{''}(\phi_{cl})}\right]\\\nonumber &=&
\frac{1}{{\cal N}^2}\sum_{n,n^{\prime}} \int d^{3}x\sqrt{g}~\int\frac{d^2k}{(2\pi)^2}\int \f{d\l~\m(\l)}{\l^2 +\n^2}(e^{-n\beta}y)(e^{-n^{\prime}\beta}y)K_{i\l}(k e^{-n\beta}y)K_{i\l}(k e^{-n^{\prime}\beta}y)e^{-i\vec{k}.(\g^n \vec{x})}e^{i\vec{k}.(\g^{n^{\prime}}\vec{x})}\\\nonumber
&=&\frac{1}{{\cal N}^2}\sum_{n,n^{\prime}} \int \f{dy}{y}~\int\frac{d^2k}{(2\pi)^2}\int \f{d\l~\m(\l)}{\l^2 +\n^2}e^{-(n+n^{\prime})\beta} K_{i\l}(k e^{-(n-n^{\prime})\beta}y)K_{i\l}(k y)(2\pi)^2\d^2(\g^n\vec{k}-\g^{n^{\prime}}\vec{k})\\\nonumber
&=&\frac{1}{{\cal N}}\sum_{n\ne 0}\f{e^{-n\beta}}{|1-e^{2\pi i n\t}|^2} \int \f{dy}{y}~\int\frac{d^2k}{(2\pi)^2}\int \f{d\l~\m(\l)}{\l^2 +\n^2} K_{i\l}(k e^{-n\beta}y)K_{i\l}(k y)(2\pi)^2\d^2(\vec{k})
\eeqa

The third line of the above equation is obtained by integrating over $\vec{x}$ and re-scaling $y$ for each term in the summation as $y\rightarrow e^{n^{\prime}\b}y$.  As before, each term in the summation over $n^{\prime}$ gives the same series in $n$, thus canceling a factor of ${\cal N}$. Further, we have discarded the zero temperature $n=0$ contribution which was computed earlier and used the delta function identity (see Appendix \ref{del})

\beqa\label{di2}
\d^2(\g^n\vec{k}-\g^{n^{\prime}}\vec{k})=\frac{e^{2n^{\prime}\b}}{|1-e^{2\pi i (n-n^{\prime})\t}|^2}\d^2(\vec{k})  
\eeqa

Next tracing the steps as in equations (\ref{ikrel})-(\ref{ikform}) and using the invariance of the expression under $n \rightarrow -n$, the trace can be written as 

\beqa
&& \frac{2}{{\cal N}}\sum_{n=1}^{\infty}\f{e^{-n\beta}}{|1-e^{2\pi i n\t}|^2} \int \f{dy}{y}~\int\frac{d^2k}{(2\pi)^2} K_{\n}(k y)I_{\n}(k e^{-n\beta}y)(2\pi)^2\d^2(\vec{k})\\\nonumber
&=& \frac{2}{{\cal N}}\sum_{n=1}^{\infty}\int \f{dy}{y}\f{e^{-n\beta}}{|1-e^{2\pi i n\t}|^2} \f{e^{-\b n\n}}{2\n}
\eeqa

where we have used,

\beqa
 K_{\n}(k y)I_{\n}(k e^{-n\beta}y) \xrightarrow[]{k \rightarrow 0}  \f{e^{-\b n\n}}{2\n} +{\cal O}(k^2)~.
\eeqa

Defining $V''(\phi_{cl})=M^2$, so that $\n=\sqrt{1+M^2}$, the finite temperature contribution to the one-loop partition function $Z^{(1)}$ is, 

\beqa
\log Z^{(1)}_{\t}&=&\f{1}{2}\int_{M^2}^{\infty}\tr\left[\frac{1}{- \Box_E+M^2}\right] dM^2\label{paV}\\
&=& \frac{1}{{\cal N}}\sum_{n=1}^{\infty}\int \f{dy}{y}\f{e^{-n\beta}}{|1-e^{2\pi i n\t}|^2} \int_{M^2}^{\infty}\f{e^{-\b n\n}}{2\n}dM^2\\
&=&\frac{1}{{\cal N}}\sum_{n=1}^{\infty}\int \f{dy}{y}\f{1}{\b n}\f{e^{-n\beta(1+\n)}}{|1-e^{2\pi i n\t}|^2} 
\eeqa

The integral over $y$ can be written as 

\beqa
\int_0^{\infty} \f{dy}{y}=\sum_{m=-\infty}^{\infty}\int_{e^{-(m+1)\b}}^{e^{-m\b}} \f{dy}{y}
={\cal N}\b
\eeqa

We note that $\b$ is the value of the $y$ integral over the fundamental region 
$e^{-\b}\le y \le 1$. Finally the one-loop partition function is,

\beqa\label{finalp3}
\log Z^{(1)}_{\t}=\sum_{n=1}^{\infty}\f{1}{n}\f{e^{-n\beta(1+\n)}}{|1-e^{2\pi i n\t}|^2} 
\eeqa

This matches with the expression derived using other methods \cite{Giombi:2008vd}-\cite{David:2009xg}. 

To make the method of images manifest consider the two-point function,   
\beqa\label{gfsteps}
&&\langle x|\left[\frac{1}{- \Box_E+V^{''}(\phi_{cl})}\right]|x^{\prime}\rangle\\\nonumber &=&
\frac{1}{{\cal N}^2}\sum_{n,n^{\prime}}\int\frac{d^2k}{(2\pi)^2}\int \f{d\l~\m(\l)}{\l^2 +\n^2}(e^{-n\beta}y)(e^{-n^{\prime}\beta}y^{\prime})K_{i\l}(k e^{-n\beta}y)K_{i\l}(k e^{-n^{\prime}\beta}y^{\prime})e^{-i\vec{k}.(\g^n \vec{x})}e^{i\vec{k}.(\g^{n^{\prime}}\vec{x}^{\prime})}\\\nonumber
&=&\frac{1}{{\cal N}}\sum_{n}\int\frac{d^2k}{(2\pi)^2}\int \f{d\l~\m(\l)}{\l^2 +\n^2}(y)(e^{-n\beta}y^{\prime})K_{i\l}(ky)K_{i\l}(k e^{-n\beta}y^{\prime})e^{-i\vec{k}.\vec{x}}e^{i\vec{k}.(\g^{n}\vec{x}^{\prime})}\\\nonumber
&=&\frac{1}{{\cal N}}\sum_{n}G(x,\g^nx^{\prime})
\eeqa

Thus,
\beqa\label{green1}
\tr\left[\frac{1}{- \Box_E+V^{''}(\phi_{cl})}\right]&=&\frac{1}{{\cal N}}\sum_{n}\int_{\mathbb{H}^3} d^3x \sqrt{g}~ G(x,\g^nx)\non
&=&\sum_{n}\int_{\mathbb{H}^3/\mathbb{Z}} d^3x \sqrt{g}~ G(x,\g^nx)
\eeqa

Following the steps below (\ref{gfsteps}), each copy of the fundamental region gives the same answer. This cancels the normalization ${\cal N}$ in the denominator in the first line of (\ref{green1}).

Before ending this section we recall that the expression for $Z^{(1)}_{(-1/\t)}$ is also equal to the partition function for a rotating BTZ black hole with parameters $(\b,\th)$ \cite{Carlip:1994gc}. This can be seen from the relation between the 
Poincar\'e coordinates $(x_1,x_2,y)$ and the Schwarzschild coordinates, $(\t,r,\phi)$. Requiring $\phi$ to be of periodicity $2\pi n$ results in the identification of $(x_1,x_2,y) \sim \g^n(x_1,x_2,y)$ with $\t$ replaced by $-1/\t$, where the action of $\g^n$ is given in equation (\ref{realt}).

\subsubsection{Thermal $AdS_{d+1}$}

The generalization of the previous computation to that of higher odd dimensional spaces ($d$ even) is straightforward. We begin by writing down the metric
 
\beqa
ds^2=\f{L^2}{y^2}\left(dy^2+\sum_{i=1}^{d/2} dz_id\bar{z_i}\right)
\eeqa

Corresponding to each $z_i$ we associate an angular transformation $\theta_i$.
Thermal $AdS_{d+1}$ is thus the quotient space with the action of $\g^n_i$ as,

\beqa\label{realtd}
\g^n_i(y,z)=(e^{-n\beta}y, e^{2\pi in\tau_i}z_i) ~~~~\mbox{where}~~~~\tau_i=\frac{1}{2\pi}(\theta_i+i\beta)
\eeqa

Writing, real coordinates $z_i=x_{i1}+ix_{i2}$, the action of each $\g^n_i$ on the real coordinates $\vec{x}_i=(x_{i1},x_{i2})$ can be written as in equation (\ref{realt}) with $\theta_i$.

The scalar wave function invariant under the transformation (\ref{realtd}) is,

\beqa
\Phi_{\vec{k},\l}(\vec{x},y)=\frac{1}{{\cal N}}\sum_{n=-\infty}^{\infty} (ke^{-n\beta}y)^{d/2}K_{i\l}(k e^{-n\beta}y)e^{-i\vec{k}_i.(\g^n_i \vec{x}_i)}
\eeqa

The normalization works out as follows,

\beqa\label{normt1}
&&\int d^{d+1}x\sqrt{g}~\Phi_{\vec{k},\l}(\vec{x},y)\Phi^{*}_{\vec{k}^{\prime},\l^{\prime}}(\vec{x},y)\\\nonumber
&=&\frac{1}{{\cal N}^2}\sum_{n,n^{\prime}} \int d^{d+1}x\sqrt{g}~(ke^{-n\beta}y)^{d/2}(k^{\prime}e^{-n^{\prime}\beta}y)^{d/2}K_{i\l}(k e^{-n\beta}y)K_{i\l^{\prime}}(k^{\prime} e^{-n^{\prime}\beta}y)e^{-i\vec{k}_i.(\g^n_i \vec{x}_i)}e^{i\vec{k}^{\prime}_i.(\g^{n^{\prime}}_i \vec{x}_i)}\\\nonumber
&=&\frac{1}{{\cal N}}\left[\d(\l-\l^{\prime})/\m(\l)\right]\sum_{n} \prod_{i=1}^{d/2}(ke^{-n\b})^2(2\pi)^2\d^2(\g^{n}_i \vec{k}_i-\vec{k}^{\prime}_i)
\eeqa

Following the steps as in $AdS_3$, the trace thus takes the following form,

\beqa
&&\tr\left[\frac{1}{- \Box_E+V^{''}(\phi_{cl})}\right]\\\nonumber &=&
\frac{1}{{\cal N}^2}\sum_{n,n^{\prime}} \int d^{3}x\sqrt{g}~\int \f{d\l~\m(\l)}{\l^2 +\n^2}\int\frac{d^dk}{(2\pi)^d}(e^{-n\beta}y)^{d/2}(e^{-n^{\prime}\beta}y)^{d/2}K_{i\l}(k e^{-n\beta}y)K_{i\l}(k e^{-n^{\prime}\beta}y)e^{-i\vec{k}_i.(\g^n_i \vec{x}_i)}e^{i\vec{k}_i.(\g^{n^{\prime}}_i\vec{x}_i)}\\\nonumber
&=&\frac{1}{{\cal N}}\sum_{n\ne 0}\int \f{dy}{y}~\int \f{d\l~\m(\l)}{\l^2 +\n^2}\int\frac{d^dk}{(2\pi)^d}  K_{i\l}(k e^{-n\beta}y)K_{i\l}(k y)\prod_{i=1}^{d/2}\f{e^{-n\beta}}{|1-e^{2\pi i n\t_i}|^2}(2\pi)^2 \d^2(\vec{k_i})\non
&=& \frac{2}{{\cal N}}\sum_{n=1}^{\infty} \int \f{dy}{y}~\int\frac{d^dk}{(2\pi)^d} K_{\n}(k y)I_{\n}(k e^{-n\beta}y)\prod_{i=1}^{d/2}\f{e^{-n\beta}}{|1-e^{2\pi i n\t_i}|^2}(2\pi)^2\d^2(\vec{k_i})\\\nonumber
&=& \frac{2}{{\cal N}}\sum_{n=1}^{\infty}\int \f{dy}{y} \f{e^{-\b n\n}}{2\n}
\prod_{i=1}^{d/2}\f{e^{-n\beta}}{|1-e^{2\pi i n\t_i}|^2}
\eeqa

This gives the following expression for the partition function

\beqa\label{zd}
\log Z=\sum_{n=1}^{\infty}\f{e^{-n\beta\n}}{n}\prod_{i=1}^{d/2}\f{e^{-n\beta}}{|1-e^{2\pi i n\t_i}|^2} 
\eeqa

For $\theta_i=0$ equation (\ref{zd}) reduces to

\beqa\label{zdt0}
\log Z=\sum_{n=1}^{\infty}\f{1}{n}\f{e^{-n\beta(d/2+\n)}}{|1-e^{-n\beta}|^d} 
\eeqa

In the case when $d$ is odd we write $z_i=x_{i1}+ix_{i2}$ as before, with $i=1,\cdots (d-1)/2$.

\beqa
ds^2=\f{L^2}{y^2}\left(dy^2+\sum_{i=1}^{(d-1)/2} dz_id\bar{z_i}+dx_d^2\right)
\eeqa

$\g^n$ acts on the coordinates as

\beqa
\g^n_i(y,z_i,x_{d})=(e^{-n\beta}y, e^{2\pi in\tau_i}z_i,e^{-n\beta}x_d)
\eeqa  

The $\g^n$ invariant solution now is, 

\beqa
\Phi_{\vec{k},\l}(\vec{x},y)=\frac{1}{{\cal N}}\sum_{n=-\infty}^{\infty} (ke^{-n\beta}y)^{d/2}K_{i\l}(k e^{-n\beta}y)e^{-i\vec{k}_i.(\g^n_i \vec{x}_i)}e^{-i{k}_d(e^{-n\b}{x}_d)}
\eeqa  

with the normalization, 

\beqa\label{normtdo}
&&\int d^{d+1}x\sqrt{g}~\Phi_{\vec{k},\l}(\vec{x},y)\Phi^{*}_{\vec{k}^{\prime},\l^{\prime}}(\vec{x},y)\\\nonumber
&=&\frac{1}{{\cal N}}\left[\d(\l-\l^{\prime})/\m(\l)\right]\sum_{n} (ke^{-n\b})(2\pi)\d(e^{-n\b}{k}_d-{k}^{\prime}_d)\prod_{i=1}^{(d-1)/2}(ke^{-n\b})^2(2\pi)^2\d^2(\g^{n}_i \vec{k}_i-\vec{k}^{\prime}_i)
\eeqa

The partition function in this case is,

\beqa\label{zdo}
\log Z=\sum_{n=1}^{\infty}\f{e^{-n\beta(1/2+\n)}}{n|1-e^{-n\beta}|}\prod_{i=1}^{(d-1)/2}\f{e^{-n\beta}}{|1-e^{2\pi i n\t_i}|^2} 
\eeqa

For $\th_i=0$, the odd $d$ case also leads to the expression (\ref{zdt0}) for the one-loop partition function, thus reproducing the result obtained in, for example \cite{Denef:2009kn},\cite{Gopakumar:2011qs},\cite{Kraus:2020nga},\cite{Keeler:2014hba}. As noted before the appearance of $\D_{+}=d/2+\n$ in (\ref{zd}), (\ref{zdo}) is due to the choice of our wave functions which are regular in the interior of $AdS$. The expressions are known to match with the CFT computations of the partition function dual to the scalar field.   
 
\section{Phases of scalar field theories}\label{phases}

In this section we study the phases of scalar theories in thermal $AdS$. We first study the theory with a single scalar field and then the $O(N)$ vector model for finite $N$ and in the large $N$ limit. The main objective is to identify the regions in the parameter space ($m^2$, $\b$) that gives various forms of the potential which correspond to the (un)broken symmetry phases. We do this numerically, setting $L=1$.

Unlike the finite temperature contribution, the zero temperature contribution to the one-loop correction is proportional to $\mbox{Vol}(\mathbb{H}^{d+1}/\mathbb{Z})$ which is divergent. We thus need the regularized volume to proceed with the analysis. Volume regularization of Euclidean $AdS$ appeared in several contexts. We first review the method using a cutoff \cite{Sen:2008vm}-\cite{Cotler:2018zff} here for $d=1,2,3$. 

To obtain the regularized volume one can use the Euclidean metric in global coordinates.
\beqa  
ds^2_{AdS_2}=d\r^2+\sinh^2\r ~d\th^2 ~~~0<\r<\infty,~0 \le \th \le 2\pi 
\eeqa
The volume with a cutoff for the radial coordinate at $\r=\r_0$, and for thermal $AdS$, taking the period of $\th$ to be $\b$,
\beqa\label{vol2}
\mbox{Vol}(\mathbb{H}^{2}/\mathbb{Z})&=&\int_0^{\b} d\th\int_0^{\r_0}\sinh\r~d\r\non
&=&\b[\cosh(\r_0)-1]=\b[\f{1}{2}(e^{\r_0}+e^{-\r_0})-1]\rightarrow {\cal V}_2=-\b
\eeqa
The ${\cal O}(e^{\r_0})$ are canceled by adding boundary counterterms at $\r=\r_0$. The resulting finite part is
${\cal V}_2=-\b$. 

Similarly,
\beqa  
ds^2_{AdS_3}=\cosh^2\r~d\t^2+d\r^2+\sinh^2\r ~d\phi^2, ~~~~~0<\r<\infty,~~0 \le \t \le 2\pi, ~~0 \le \phi \le 2\pi 
\eeqa
\beqa\label{vol3}
\mbox{Vol}(\mathbb{H}^{3}/\mathbb{Z})&=&\int_0^{2\p}d\phi\int_0^{\b} d\t \int_0^{\r_0}\cosh\r\sinh\r~d\r\non
&=&\f{\pi\b}{2}[\cosh(2\r_0)-1]=\pi\b[\f{1}{2}(e^{2\r_0}+e^{-2\r_0})-1]\rightarrow {\cal V}_3=-\f{\pi\b}{2}
\eeqa
Finally,
\beqa  
ds^2_{AdS_4}=d\r^2+\sinh^2\r ~d\O_3^2, ~~~~~0<\r<\infty 
\eeqa
\beqa\label{vol4}
\mbox{Vol}(\mathbb{H}^{4}/\mathbb{Z})&=&\mbox{Vol}(\O_{3})\f{\b}{2\pi}\int_0^{\r_0}\sinh^3\r~d\r\non
&=&\pi\b[\f{1}{3}\cosh^3(\r_0)-\cosh(\r_0)+\f{2}{3}]\non
&=&\pi\b[\f{1}{24}e^{3\r_0}-\f{3}{8}e^{-2\r_0}+\f{2}{3}+{\cal O}(e^{-\r_0})]\rightarrow {\cal V}_4=\f{2\pi\b}{3}
\eeqa

Dimensional regularization was used in \cite{Graham:1999jg},\cite{Diaz:2007an}. The regulatized volume of $\mathbb{H}^{d+1}/\mathbb{Z}$ is given by ${\cal V}_{d+1}={\cal V}(\mathbb{H}^{d+1})\b/(2\pi)$ where 
 
 \beqa\label{dimregv}
{\cal V}(\mathbb{H}^{d+1})&=&\f{(-\pi)^{d/2}}{\G((d+2)/2)}\left[\psi(1+d/2)-\log \pi\right] ~~~~~\mbox{for}~~~~~\mbox{even}~~~d~~~\non
&=&  (-1)^{(d+1)/2}\f{\pi^{(d+2)/2}}{\G((d+2)/2)} ~~~~~~~~~~~~~~~~~~\mbox{for}~~~~~\mbox{odd}~~~d~~~
\eeqa

We shall see that the sign of the regularized volume leads to qualitative differences between the phases of scalar theories, using the following perturbative analysis. The regularized volume for $d=2$ (thermal $AdS_3$) differ both in magnitude and in sign for the two methods while for $d=1,3$ they are same. As mentioned in the introduction we could thus use any of the regularization schemes for the volume in our analysis of phases for $AdS_3$ each giving a different qualitative behavior. In the following we  present the details for both the cases.

\subsection{Single scalar}\label{single}

Consider the following Lagrangian for the $\phi^{4}$ theory
\begin{equation}
\label{ch4e1}
\mathcal{L}_E=\frac{1}{2} ( \partial ^{}_{\mu} \phi)^{2} + \frac{1}{2} m^{2} \phi^2 + \frac{\lambda}{4!}  \phi ^{4}
\end{equation}
The effective potential for the $\phi^{4}$ at finite temperature can thus be written as

\beqa
V^{}_{eff}(\phi^{}_{cl})&=&V(\phi_{cl})-\f{1}{{\cal V}_{d+1}}[\log Z^{(1)}+\log Z^{(1)}_{\t}]
\eeqa

We shall consider the theory on $AdS_3$. Thus setting $d=2$ in (\ref{trzt}) gives

\beqa\label{trzt3}
\f{1}{2}\mbox{tr}\left[\f{1}{-\square_E+V^{''}(\phi_{cl})}\right]=-{\cal V}_3\f{\sqrt{1+M^2}}{8\pi}
\eeqa

$\n$ and the effective mass, $M$ are related as

\begin{equation}
\label{ch4e3}
\n=\sqrt{1+M^2}=\left[1+\f{\lambda}{2}\phi^{2}_{cl}+m^{2}\right]^{1/2}.
\end{equation}

$\log Z^{(1)}$ is obtained from (\ref{trzt3}) after integrating over $M^2$ as in equation (\ref{paV}). We discard an infinite constant arising from the integral on the RHS of (\ref{paV}). $\log Z^{(1)}_{\t}$ is  given in equation (\ref{finalp3}). 
We further set the angular potential $\theta=0$ and insert ${\cal V}_{3}=-\beta \pi/2$ from (\ref{vol3}). The phase plot corresponding to the positive renormalized volume as in (\ref{dimregv}) is given in the next section. 

The complete expression for the one-loop corrected effective potential is,

\begin{eqnarray}\label{ads3finitet1}
V^{}_{eff}(\phi^{}_{cl})=\frac{1}{2} m^{2}\phi^{2}_{cl}+\frac{\lambda}{4!}\phi^{4}_{cl} -\frac{\n^3}{12\pi} +\f{2}{\pi  \b}\sum _{n=1}^{\infty} \frac{ e^{-\b n (1+\n)}}{ n (1-e^{-\b n})^2}
\end{eqnarray}

As a side remark, we note that the one-loop zero temperature contribution to the effective potential for $AdS_3$ resulting from equation (\ref{trzt}) is finite. Thus no counterterm is added in (\ref{ads3finitet1}) and accordingly no renormalization condition is imposed. The mass $m$, that appears in equation (\ref{ads3finitet1}) is the renormalized mass. The same observation holds for the analysis on $AdS_3$ in sections \ref{ads3fn} and \ref{ads3n}.

\begin{figure}[H] 
\begin{center} 
  \begin{minipage}{0.5\textwidth}%   
   \begin{subfigure}[b]{0.8\linewidth}
    \centering
    \includegraphics[width=1.1\linewidth]{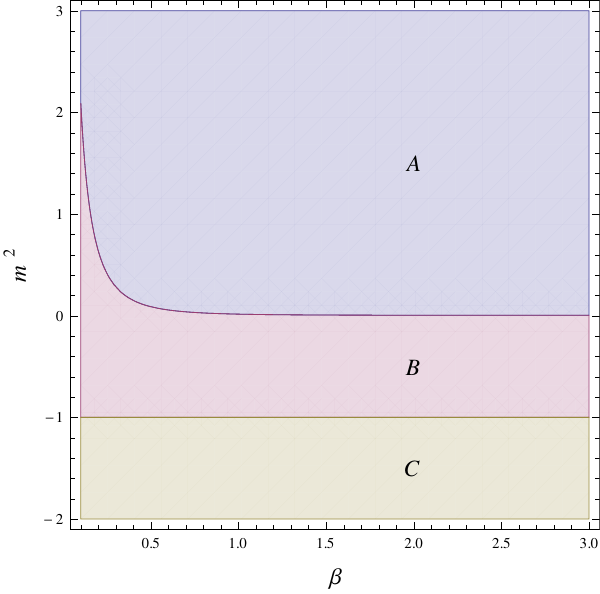} 
    \caption{Phases on the $\b-m^2$ plane} 
    \label{singleregions} 
    \vspace{1ex}
  \end{subfigure}
  \end{minipage}%
  % \hspace{-1cm}
  \begin{minipage}{0.7\textwidth}%
  \begin{minipage}{0.7\textwidth}%
  \begin{subfigure}[b]{0.5\linewidth}
    \centering
    \includegraphics[width=1.35\linewidth]{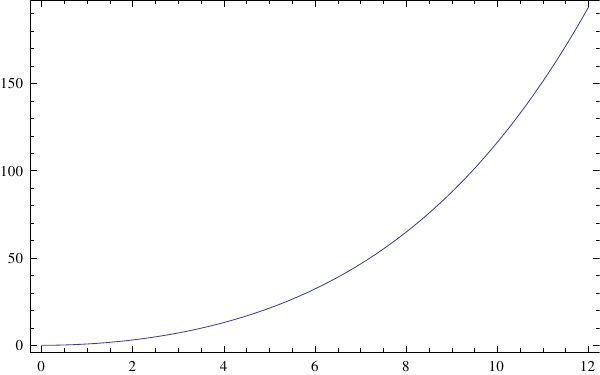} 
    \caption{$A$: $m^2=1.5$, $\b=2.0$} 
    \label{singlea} 
    \vspace{1ex}
  \end{subfigure}
   \end{minipage}%
   \hspace{-2em}
   \begin{subfigure}[b]{0.5\linewidth}
    \centering
    \includegraphics[width=0.9\linewidth]{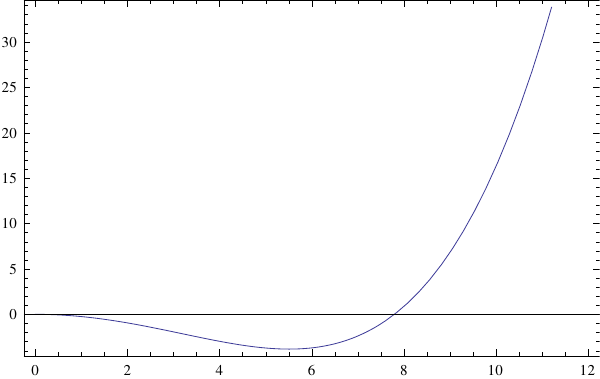} 
    \caption{$B$: $m^2=-0.5$, $\b=2.0$} 
    \label{singleb} 
    \vspace{1ex}
  \end{subfigure}
   \end{minipage}
 % \begin{subfigure}[b]{0.5\linewidth}
  %  \centering
   % \includegraphics[width=0.75\linewidth]{C2.eps} 
    %\caption*{$C$: $m^2=-1.0$,$\b=2.0$} 
    %\label{c} 
  %\end{subfigure}%%
  \caption{Phases and potentials for $\l=0.1$. }
  \label{singleall} 
  \end{center} 
\end{figure}

We are thus adopting the minimal subtraction scheme in writing down 
effective potentials for $AdS_3$. An alternate renormalization scheme used in the analysis on $AdS_2$ and $AdS_4$ in the later sections, utilizes the mass renormalization condition as in equations (\ref{sads2rc}) and (\ref{renorcads4}). The qualitative nature of the phase plots are expected to remain the same in both the schemes.

The saddle point equation from (\ref{ads3finitet1}) is

\begin{equation}
\label{ch4e4}
0=\frac{dV^{}_{eff}}{d\phi^{}_{cl}}= -\sum _{n=1}^{\infty}\f{\lambda\phi^{}_{cl}}{\pi\n}\frac{ e^{-\beta n \left(1+\n\right)}}{  \left(1-e^{-\beta n}\right)^2 }-\frac{\lambda \phi^{}_{cl} \n}{8 \pi }+\frac{\lambda \phi^{3}_{cl}}{6}+m^{2}\phi^{}_{cl}.
\end{equation}

 Figure \ref{singleregions} shows the regions in the $\b-m^2$ parameter space corresponding to the various forms of the potentials shown in Figures \ref{singlea} and \ref{singleb} . These regions are obtained numerically for $n=100$. In the region $A$ the theory has only one minimum at $\phi_{cl}=0$. The region $B$ corresponds to the theory having a minimum at $\phi_{cl}\ne 0$ and a maximum at $\phi_{cl}=0$. On the boundary separating $A$ and $B$ both these extrema coincide and the line at zero temperature approaches a $m^2$ value which is a solution of $m^2=\l/(8\pi)\sqrt{1+m^2}$. Since the region $B$ is bounded below by the Breitenlohner-Freedman (BF) bound $m^2=-1$ both the symmetry breaking $\phi_{cl}\ne 0$ and the symmetry preserving $\phi_{cl}=0$ coexist, unlike the theory in flat space. The region $C$ is below the BF bound and hence unstable\footnote{The effective potential in the alternate scheme including a mass counterterm in (\ref{ads3finitet1}) and imposing (\ref{sads2rc}) gives $V^{}_{eff}(\phi^{}_{cl})=\frac{1}{2} (m^{2}+\f{\l}{8\pi}\sqrt{1+m^2})\phi^{2}_{cl}+\frac{\lambda}{4!}\phi^{4}_{cl} -\frac{\n^3}{12\pi} +\f{2}{\pi  \b}\sum _{n=1}^{\infty} \frac{ e^{-\b n (1+\n)}}{ n (1-e^{-\b n})^2}$. The qualitative nature of the phase plot is same as that of Figure \ref{singleall}. The boundary separating regions $A$ and $B$ now asymptotes to $m^2=0$ at zero temperature.}. 

We now consider the theory on $AdS_2$. The zero temperature trace (\ref{trzt}) is divergent for $d=1$. We thus expand 

\beqa\label{expads2}
\f{\m^{-\e}}{{2\cal V}_{d+1}}\mbox{tr} \frac{1}{-\square_E+M^{2}}=\frac{1}{4 \pi \e}+\frac{1}{8\pi}\left[-2 \psi^{(0)}\left(\n+\frac{1}{2}\right)-\gamma +\log (4\pi )-\log(\m^2)\right]+{\cal O}\left(d-1\right)
\eeqa

where $\gamma$ is the Euler-Mascheroni constant, $\psi^{(m)}(x)=d^{m+1} \log\Gamma(x)/dx^{m+1}$ is the Polygamma function, 
$\e=1-d$ and ${\m}$ is a parameter having mass dimension which is introduced as usual in dimensional regularization to compensate for the dimensions of the parameters. The divergence appears as a pole $1/\e$ in the above expression. 

The renormalized effective potential is obtained by including a counterterm  so that

\begin{eqnarray}\label{ads2effpot}
V^{}_{eff}(\phi^{}_{cl})=\frac{1}{2} m^{2}\phi^{2}_{cl}+\frac{\lambda}{4!}\phi^{4}_{cl} +M^2\d m^2-\f{1}{{\cal V}_{d+1}}(\log Z^{(1)}+\log Z^{(1)}_{\t}).
\end{eqnarray}

The counterterm $\d m^2$ is obtained by imposing the following renormalization condition on zero temperature effective potential $V_{eff}^0$.

\beqa\label{sads2rc}
\left.\f{\pa^2}{\pa \phi_{cl}^2}V^{0}_{eff}(\phi^{}_{cl})\right|_{\phi_{cl}=0}=m^2
\eeqa

where from (\ref{ads2effpot}), $V^{0}_{eff}$ is 
\beqa
V^{0}_{eff}=\frac{1}{2} m^{2}\phi^{2}_{cl}+\frac{\lambda}{4!}\phi^{4}_{cl} +M^2\d m^2-\f{{\m}^{-\e}}{2{\cal V}_{d+1}}\int_{M^2}^{\infty}\tr \left[\frac{1}{-\square_E+M^2}\right]dM^2.
\eeqa

With $M^2$ defined in (\ref{ch4e3}), this gives

\beqa
\d m^2&=&-\f{{\m}^{-\e}}{2{\cal V}_{d+1}}\tr \left[\frac{1}{-\square_E+M^2}\right]_{\phi_{cl}=0}\non
&=&-\f{1}{4\pi\e}-\f{1}{8\pi}\left[-2 \psi^{(0)}\left(\n+\frac{1}{2}\right)-\g +\log (4\pi) -\log({\m}^2)\right]+{\cal O}(\e).
\eeqa

The integral over $M^2$ can be written as,

\beqa
=-\frac{1}{4\pi}\int^{M^{2}}_{0} dM^2 \left[\psi^{(0)}\left(\n+\frac{1}{2}\right)\right]+~~\mbox{infinite constant}.
\eeqa

We shall discard the infinite constant as before in our analysis. Next including the finite temperature contribution from (\ref{zdt0}) and putting ${\cal V}^{}_{2}=-\beta$ from (\ref{vol2}), the effective potential at finite temperature becomes

\beqa
V^{}_{eff}(\phi^{}_{cl})&=&\frac{1}{2} m^{2}\phi^{2}_{cl}+\frac{\lambda}{4!}\phi^{4}_{cl}
+\frac{1}{2}\int^{M^{2}}_{0} dM^2 \left[\frac{1}{2\pi}\left(\psi^{(0)}\left(\sqrt{\f{1}{4}+m^2}+\frac{1}{2}\right)-\psi^{(0)}\left(\n+\frac{1}{2}\right)\right)\right]\non
&+&\f{1}{\b}\sum^{\infty}_{n=1}\frac{1}{n}\frac{e^{-n\beta\left(\frac{1}{2}+\sqrt{\frac{1}{4}+M^2}\right)}}{|1-e^{-n\beta}|}
\eeqa

The phase diagram for the single scalar theory on $AdS_2$ is similar to that of the theory on $AdS_3$ (with negative renormalized volume) as shown in Figure \ref{singleregions}. The renormalization condition (\ref{sads2rc}) implies that on the boundary separating the regions $A$ and $B$,
the $m^2$ value approaches zero for large $\b$. 

The symmetry breaking phase persists at high temperatures for large enough positive mass values, unlike that in the flat space where the symmetry is restored for this model at high temperatures. Symmetry restoration in flat space however does not occur for all models, see for example \cite{Weinberg:1974hy}. In a recent work \cite{Chai:2020zgq} the authors have addressed this issue.

\subsection{$O(N)$ vector model}\label{onmodel}

The Lagrangian for the $O(N)$ vector model is given by

\beqa\label{onlag}
\mathcal{L}_E=\frac{1}{2} ( \partial ^{}_{\mu} \phi^i)^{2} + \frac{1}{2} m^{2} (\phi^i)^2 + \frac{\lambda}{4} \left[(\phi^i)^2\right ]^{2}~~~~~\mbox{where}~~~i=1,\cdots,N~.
\eeqa

Expanding about the classical field as $\phi^i=\phi_{cl}^i+\eta^i$ and setting $\phi^{i}_{cl}=(0,0,\cdots,0,\phi^{}_{cl})$,  gives a modified Klein-Gordon operator $[-\square^{}_{E}+M^{2}_{i}]$ with
\begin{equation}
M^{2}_{i}=
\begin{cases}
\lambda\phi^{2}_{cl}+m^{2}, & \text{for}\ \eta^{1}\cdots\eta^{N-1} \\
3\lambda\phi^{2}_{cl}+m^{2}, & \text{for}\ \eta^{N}
\end{cases} 
\end{equation}
The effective potential thus becomes
\begin{equation}
V^{}_{eff}(\phi^{}_{cl})=V(\phi^{}_{cl})+\frac{1}{2{\cal V}^{}_{d+1}}\bigg[{(N-1)}\mbox{tr} \log[-\square^{}_{E}+M_1^{2}]+\mbox{tr} \log[-\square^{}_{E}+M_2^{2}]\bigg]
\end{equation}

The modification here from that of the flat-space case \cite{Peskin:1995ev} is encoded in the traces.

\subsubsection{$AdS^{}_{2,3}$}\label{ads3fn}

Similar to the case of the single scalar theory, where the trace is given by (\ref{trzt3}), the leading contribution to the effective potential for the $O(N)$ vector model including the expression for the partition function (\ref{finalp3}) can be written for $AdS_2$ and $AdS_3$. Here we give the relevant expressions for $AdS_3$.

\begin{eqnarray}
V^{}_{eff}(\phi^{}_{cl})&=&\frac{1}{2}m^{2}\phi^{2}_{cl}+\frac{\lambda}{4}\phi_{cl}^{4}-
\frac{1}{12\pi}\big[(N-1)\n(M_1^{2})^{3} +\n(M_2^{2})^{3}\big] \non
&-&\f{(N-1)}{{\cal V}_3}\sum^{\infty}_{n=1}\frac{1}{n}\frac{e^{-n\beta\left(1+\n(M_1^2)\right)}}{|1-e^{2\pi i n\t}|^{2}}-\f{1}{{\cal V}_3}\sum^{\infty}_{n=1}\frac{1}{n}\frac{e^{-n\beta\left(1+\n(M_2^2)\right)}}{|1-e^{2\pi i n\t}|^{2}}
\end{eqnarray}

where $\n(M_i^2)=\sqrt{1+M_i^2}$. Extremizing the potential gives,

\begin{eqnarray}
0=\frac{\partial V}{\partial \phi^{}_{cl}}=\l \phi^{3}_{cl}+m^{2}\phi^{}_{cl}-\frac{ (N-1)\l \phi^{}_{cl} \n(M_1^{2})}{4 \pi }-\frac{3\l\phi^{}_{cl} \n(M_2^2)}{4 \pi }\non
+\sum ^{\infty}_{n=1} \frac{\b (N-1)\l\phi^{}_{cl} e^{-\b n \left(1+\n(M_1^{2})\right)}}{{\cal V}_3   \left(1-e^{-\b n}\right)^2 \n(M_1^{2})}
+\sum ^{\infty}_{n=1} \frac{3\b \l \phi^{}_{cl} e^{-\b n \left(1+\n(M_2^{2})\right)}}{{\cal V}_3  \left(1-e^{-\b n}\right)^2 \n(M_2^{2})}
\end{eqnarray}

where the angular potential has been set as $\theta=0$. The $\b-m^2$ phase plots for $AdS_2$ and $AdS_3$ with negative renormalized volume are qualitatively similar to that in Figure \ref{singleregions}.  Unlike flat space, in two dimensions the symmetry broken phase exists for $AdS_2$. This was noted for the large $N$ vector model in \cite{Carmi:2018qzm}, \cite{Inami:1985dj} which we shall study in section \ref{ads2ln}. As already seen for the single scalar theory discussed above, another contrasting feature from that of flat space is that, one gets a broken symmetry phase at high temperatures.

\begin{figure}[H] 
\begin{center} 
  \begin{minipage}{0.5\textwidth}%   
   \begin{subfigure}[b]{0.8\linewidth}
    \centering
    \includegraphics[width=1.1\linewidth]{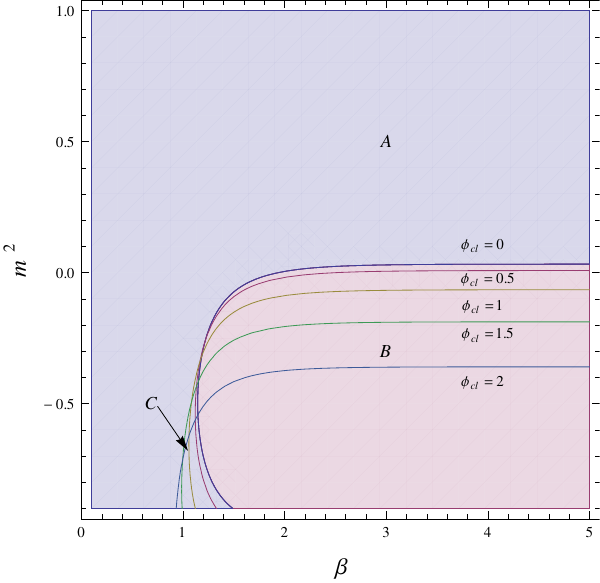} 
    %\caption{Phases on the $\b-m^2$ plane} 
    \label{OnAds4region} 
    \vspace{1ex}
  \end{subfigure}
  \end{minipage}%%%
  \begin{minipage}{0.5\textwidth}%
  \begin{subfigure}[b]{0.5\linewidth}
    \centering
    \includegraphics[width=0.95\linewidth]{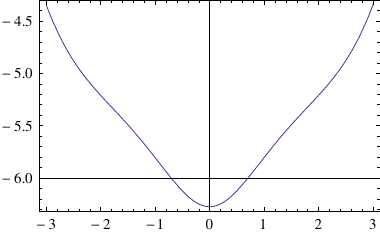} 
    \caption*{$A$: $m^2=-0.7$, $\b=0.83$} 
    %\label{a} 
    \vspace{1ex}
  \end{subfigure}%
  %\hspace{-10em}
  \begin{subfigure}[b]{0.5\linewidth}
    \centering
    \includegraphics[width=0.95\linewidth]{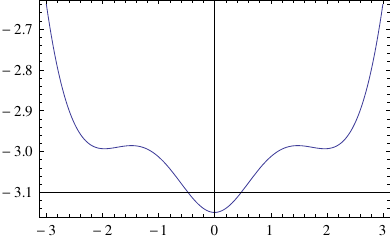} 
    \caption*{$C$: $m^2=-0.7$, $\b=1$} 
    %\label{b} 
    \vspace{1ex}
  \end{subfigure} 
  \begin{subfigure}[b]{0.5\linewidth}
    \centering
    \includegraphics[width=0.95\linewidth]{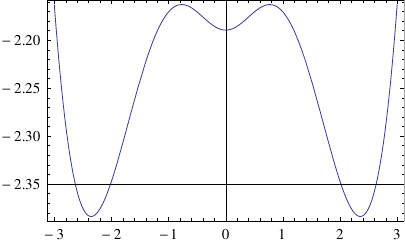} 
    \caption*{$C$: $m^2=-0.7$, $\b=1.1$} 
    %\label{c} 
  \end{subfigure}%%
  %\hspace{-5em}
  \begin{subfigure}[b]{0.5\linewidth}
    \centering
    \includegraphics[width=0.95\linewidth]{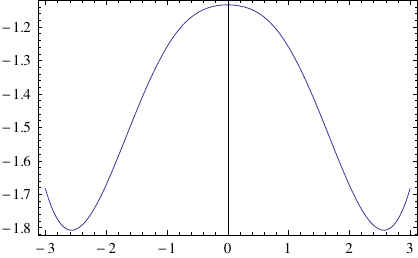} 
   \caption*{$B$: $m^2=-0.7$, $\b=1.3$} 
    %\label{d} 
  \end{subfigure} 
 % \label{fig7} 
\end{minipage}
  \caption{Phases and potentials for $\l=0.1$, $N=2$ and $n=100$. The potential plots are shown for regions $A$, $B$ and $C$. }
  \label{Onads3p} 
  \end{center} 
\end{figure}

For the positive regularized volume of thermal $AdS_3$ as in (\ref{dimregv}) the phase plot is given in Figure \ref{Onads3p}. On the boundary separating $A$ and $B$, two extrema at $\phi_{cl}=0$ coincide. The contours for the minima of the potential that appear for various values $\phi_{cl}$ intersect in region $C$ which gives rise to the existence of an additional extremum as compared to the previous cases. The effective mass-squared being above the BF bound all the phases coexist in region $C$. The preferred symmetric phase at $\phi_{cl}=0$ gives way to the broken symmetry phase as $\b$ is increased. Thus unlike the previous case of negative renormalized volume, at high temperatures we see that the symmetry is restored. In region $A$ the broken symmetry phase is always preferred. For $N=1$ the phase plot is qualitatively similar as described above.

\subsubsection{$AdS_4$}

Expanding the expression for the trace at zero temperature around $d=3$ and substituting $L=1$, gives
\begin{eqnarray}\label{zads4exp}
\frac{\m^{-\e}}{{\cal V}^{}_{d+1}}\mbox{tr} \frac{1}{-\square+M_i^{2}}=\frac{(2+M_i^2)}{16\pi^2}\left[-\frac{2}{\e}-1+\gamma-\log(4\pi)+\psi^{(0)}\left(\n(M_i^2)-\f{1}{2}\right)+\psi^{(0)}\left(\n(M_i^2)+\f{3}{2}\right)+\log(\m^2)\right]\non
\end{eqnarray}

where $\n(M_i^2)=\sqrt{9/4+M_i^2}$ and $\e=3-d$. The effective potential is renormalized by adding counterterms corresponding to $m^2$ and $\l$ 
so that,

\begin{eqnarray}\label{effnct4}
V_{eff}(\phi_{cl}) =\frac{1}{2}m^{2}\phi^{2}_{cl}+\frac{\lambda}{4}\phi_{cl}^{4}-\f{1}{{\cal V}_{d+1}}(\log Z^{(1)}+\log Z^{(1)}_{\b})
+ \f{\d m^2}{2} \phi_{cl}^2+\f{\d\l}{4}\phi_{cl}^4
\end{eqnarray}

where the zero temperature contribution is,

\beqa
-\f{1}{{\cal V}_{d+1}}\log Z^{(1)}=\f{\l}{{2\cal V}_{d+1}}\int_0^{\phi_{cl}^2}\left[(N-1)\mbox{tr} \frac{1}{-\square+M^{2}_1}+3\mbox{tr} \frac{1}{-\square+M^{2}_2}\right]d\phi_{cl}^2 .
\eeqa

To renormalize we set the following renormalization conditions (at zero temperature),

\beqa\label{renorcads4}
\left.\frac{\partial }{\partial \phi_{cl}^{2}}V^{0}_{eff}(\phi_{cl})\right|_{\phi_{cl}=0}=\f{m^2}{2}
~~~~~~~
\left.\frac{\partial^2 }{\partial (\phi_{cl}^2)^2}V^{0}_{eff}(\phi_{cl})\right|_{\phi_{cl}=0}=\f{\l}{2}
\eeqa

which give

\beqa
\d m^2=-\l(N+2)\f{\m^{-\e}}{{\cal V}_{d+1}}\mbox{tr} \frac{1}{-\square+m^{2}}~~~\mbox{and}~~~\d \l=-\l^2(N+8)\f{\m^{-\e}}{{\cal V}_{d+1}}\f{\partial}{\partial m^2}\mbox{tr} \frac{1}{-\square+m^{2}}
\eeqa

The renormalized effective potential at zero temperature, removing (an infinite) constant is thus

\begin{eqnarray}\label{effnct4ren}
V_{eff}(\phi_{cl}) &=&\frac{1}{2}m^{2}\phi^{2}_{cl}+\frac{\lambda}{4}\phi_{cl}^{4}+ \f{\l}{{2\cal V}_{d+1}}\int_0^{\phi_{cl}^2}\left[(N-1)\mbox{tr} \frac{1}{-\square+M^{2}_1}+3\mbox{tr} \frac{1}{-\square+M^{2}_2}\right]_{\mbox{ren}}d\phi_{cl}^2\non
&-&\f{(N+8)\l^2}{128\pi^2}\f{(2+m^2)}{\n(m^2)}\left[\psi^{(1)}\left(\n(m^2)-\f{1}{2}\right)+\psi^{(1)}\left(\n(m^2)+\f{3}{2}\right)\right]\phi_{cl}^4
\end{eqnarray}

where

\beqa
&&\f{1}{{\cal V}_{d+1}}\left[(N-1)\mbox{tr} \frac{1}{-\square+M^{2}_1}+3\mbox{tr} \frac{1}{-\square+M^{2}_2}\right]_{\mbox{ren}}\non
&=&
(N-1)\frac{(2+M_1^2)}{16\pi^2}\left[\psi^{(0)}\left(\n(M_1^2)-\f{1}{2}\right)+\psi^{(0)}\left(\n(M_1^2)+\f{3}{2}\right)-\psi^{(0)}\left(\n(m^2)-\f{1}{2}\right)-\psi^{(0)}\left(\n(m^2)+\f{3}{2}\right)\right]\non
&+&3\frac{(2+M_2^2)}{16\pi^2}\left[\psi^{(0)}\left(\n(M_2^2)-\f{1}{2}\right)+\psi^{(0)}\left(\n(M_2^2)+\f{3}{2}\right)-\psi^{(0)}\left(\n(m^2)-\f{1}{2}\right)-\psi^{(0)}\left(\n(m^2)+\f{3}{2}\right)\right]
\eeqa

Including the finite temperature contribution and extremizing the potential,

\begin{eqnarray}
0=\frac{\partial V}{\partial \phi^{}_{cl}}&=&\l \phi^{3}_{cl}+m^{2}\phi^{}_{cl}+\f{\l\phi_{cl}}{{\cal V}_{d+1}}\left[(N-1)\mbox{tr} \frac{1}{-\square+M^{2}_1}+3\mbox{tr} \frac{1}{-\square+M^{2}_2}\right]_{\mbox{ren}}\non
&-&\f{(N+8)\l^2}{32\pi^2}\f{(2+m^2)}{\n(m^2)}\left[\psi^{(1)}\left(\n(m^2)-\f{1}{2}\right)+\psi^{(1)}\left(\n(m^2)+\f{3}{2}\right)\right]\phi_{cl}^3\non
&+&\sum ^{\infty}_{n=1} \frac{3 (N-1)\l\phi^{}_{cl} e^{-\b n \left(3/2+\n(M_1^2)\right)}}{2\pi  \left(1-e^{-\b n}\right)^3  \n(M_1^2)}
+\sum ^{\infty}_{n=1} \frac{9 \l \phi^{}_{cl} e^{-\b n \left(3/2+ \n(M_2^2)\right)}}{2\pi  \left(1-e^{-\b n}\right)^3 \n(M_2^2)}.
\end{eqnarray}

\begin{figure}[H] 
\begin{center} 
  \begin{minipage}{0.5\textwidth}%   
   \begin{subfigure}[b]{0.8\linewidth}
    \centering
    \includegraphics[width=1.1\linewidth]{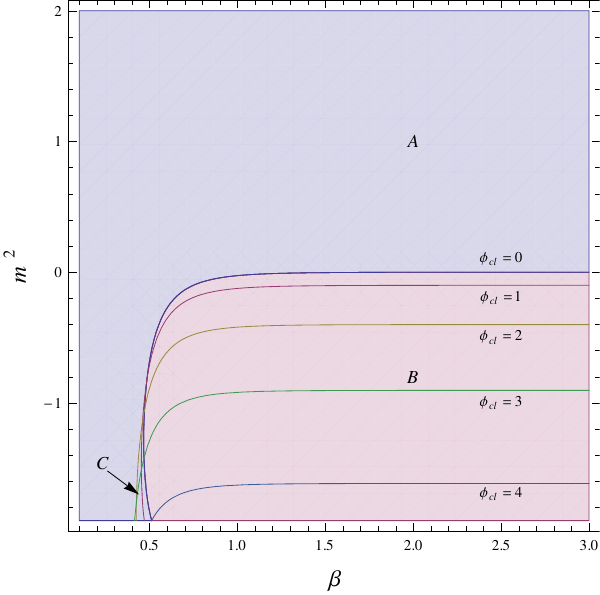} 
    %\caption{Phases on the $\b-m^2$ plane} 
    \label{OnAds4region} 
    \vspace{1ex}
  \end{subfigure}
  \end{minipage}%%%
  \begin{minipage}{0.5\textwidth}%
  \begin{subfigure}[b]{0.5\linewidth}
    \centering
    \includegraphics[width=0.95\linewidth]{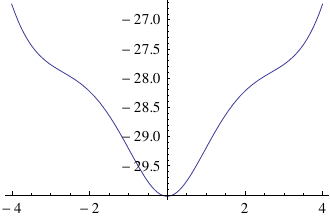} 
    \caption*{$A$: $m^2=-1.86$, $\b=0.4$} 
    %\label{a} 
    \vspace{1ex}
  \end{subfigure}%
  %\hspace{-10em}
  \begin{subfigure}[b]{0.5\linewidth}
    \centering
    \includegraphics[width=0.95\linewidth]{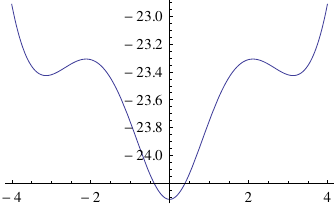} 
    \caption*{$C$: $m^2=-1.86$, $\b=0.42$} 
    %\label{b} 
    \vspace{1ex}
  \end{subfigure} 
  \begin{subfigure}[b]{0.5\linewidth}
    \centering
    \includegraphics[width=0.95\linewidth]{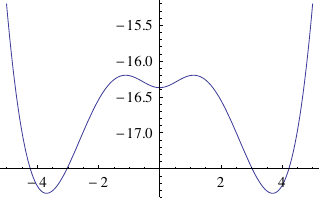} 
    \caption*{$C$: $m^2=-1.86$, $\b=0.46$} 
    %\label{c} 
  \end{subfigure}%%
  %\hspace{-5em}
  \begin{subfigure}[b]{0.5\linewidth}
    \centering
    \includegraphics[width=0.95\linewidth]{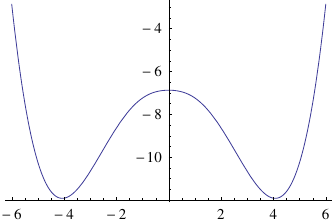} 
   \caption*{$B$: $m^2=-1.86$, $\b=0.56$} 
    %\label{d} 
  \end{subfigure} 
 % \label{fig7} 
\end{minipage}
  \caption{Phases and potentials for $\l=0.1$, $N=2$ and $n=20$. The potential plots are shown for regions $A$, $B$ and $C$. }
  \label{Onads4} 
  \end{center} 
\end{figure}

The phase plot (Figure \ref{Onads4}) in this case is qualitatively different from that of $AdS_2$ and $AdS_3$ with negative renormalized volume but similar to that of $AdS_3$ with positive renormalized volume. Figure \ref{Onads4} shows the phase plots and the potentials corresponding to the various regions for $\l=0.1$ and $N=2$.

\subsection{Large $N$}\label{largen}

We begin with some basics using the notation as in \cite{Carmi:2018qzm}. The Lagrangian for the $O(N)$ vector model is given by
(\ref{onlag}).

For organizing the perturbation theory in $1/N$ one re-scales the coupling $\l\rightarrow \l/N$ and introduces an auxiliary field $\s$ so that the Lagrangian is

\begin{equation}
\mathcal{L}=\frac{1}{2}(\partial_{\m}\phi^{i})^{2}+\frac{m^{2}}{2}(\phi^{i})^{2}-\frac{1}{2\lambda}\sigma^{2}+\frac{1}{\sqrt{N}}\sigma(\phi^{i})^{2}
\end{equation}

Expanding the fields as $\phi^i(x)=\sqrt{N}\phi_{cl}^i+\d\phi^i(x)$ and $\s(x)=\sqrt{N}\s_{cl}+\d\s(x)$, and then performing the integral over the fluctuations $\d\phi^i(x)$ one gets the following effective potential to the leading order in $1/N$.

\begin{equation}
V_{eff}(\phi_{cl}^i,\s_{cl})=N\left[\frac{M^{2}}{2}(\phi^{i}_{cl})^{2}-\frac{(M^{2}-m^2)^2}{8\l}+\frac{1}{2} \mbox{tr} \log \left(-\square^{}_{E} +M^2\right)\right]
\end{equation}

where $M^2=m^2+2\s_{cl}$. Writing the trace in the above equation by separating the contributions from the zero and finite temperature, 

\begin{equation}\label{effln}
V_{eff}(\phi_{cl}^i,\s_{cl})=N\left[\frac{M^{2}}{2}(\phi^{i}_{cl})^{2}-\frac{(M^{2}-m^2)^2}{8\l}-\f{1}{{\cal V}_{d+1}}(\log Z^{(1)}+\log Z^{(1)}_{\t})\right].
\end{equation}

In the following subsections we shall study the phases in various dimensions.

\subsubsection{$AdS_3$}\label{ads3n}

As in the case of the single scalar theory, with the trace given by (\ref{trzt3}), the leading contribution to the effective potential  for the large $N$ theory (\ref{effln}) at zero temperature, after removing an infinite constant is,
 
\begin{equation}
\label{e122}
\frac{V^0_{eff}(M^{2},\phi^{i}_{cl})}{N}=-\frac{\left(M^{2}-m^{2}\right)^{2}}{8\lambda}+\frac{1}{2}(\phi^{i}_{cl})^{2}M^{2}-\frac{(1+M^{2})^{\frac{3}{2}}}{12\pi}
\end{equation}

Including the expression for the partition function (\ref{finalp3}), 

\begin{eqnarray}
\frac{V^{}_{eff}(M^{2},\phi^{i}_{cl})}{N}=-\frac{\left(M^{2}-m^{2}\right)^{2}}{8\lambda}+\frac{1}{2}(\phi^{i}_{cl})^{2}M^{2}-\frac{(1+M^{2})^{\frac{3}{2}}}{12\pi} -\f{1}{{\cal V}_3}\sum^{\infty}_{n=1}\frac{1}{n}\frac{e^{-n\beta\left(1+\sqrt{1+M^2}\right)}}{|1-e^{2\pi i n\t}|^{2}}
\end{eqnarray}

The saddle point equation is
\begin{eqnarray}\label{saddle3}
0=\frac{1}{N}\frac{\partial V_{eff}}{\partial M^{2}}=\frac{m^{2}-M^{2}}{4\lambda}+\frac{(\phi^{i}_{cl})^{2}}{2}-\frac{\sqrt{1+M^{2}}}{8\pi}+\f{1}{{\cal V}_3}\sum^{\infty}_{n=1}\frac{\b e^{-n\beta\left(1+\sqrt{1+M^2}\right)}}{2|1-e^{2\pi i n \t}|^{2}\sqrt{1+M^{2}}}
\end{eqnarray}

\begin{figure}[H]
\centering
\begin{subfigure}[b]{0.4\linewidth}
\includegraphics[width=7cm,height=5cm]{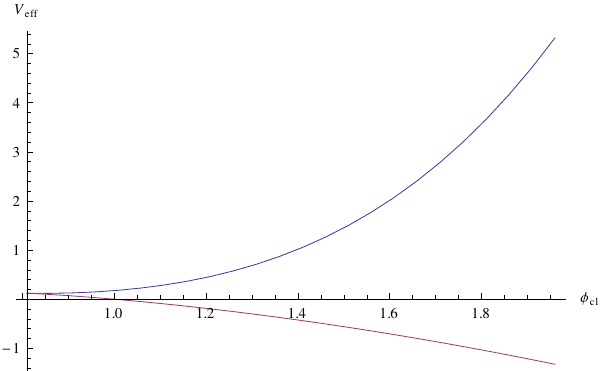}
\caption{Effective potentials corresponding to the two roots of the saddle point equation for $\lambda=1,  m^2=-0.8$ , $\beta=1$ and $n=10$.}
\label{potential}
\end{subfigure}% 
\hspace{2mm}
\begin{subfigure}[b]{0.4\linewidth}
\centering
\includegraphics[width=7cm,height=5cm]{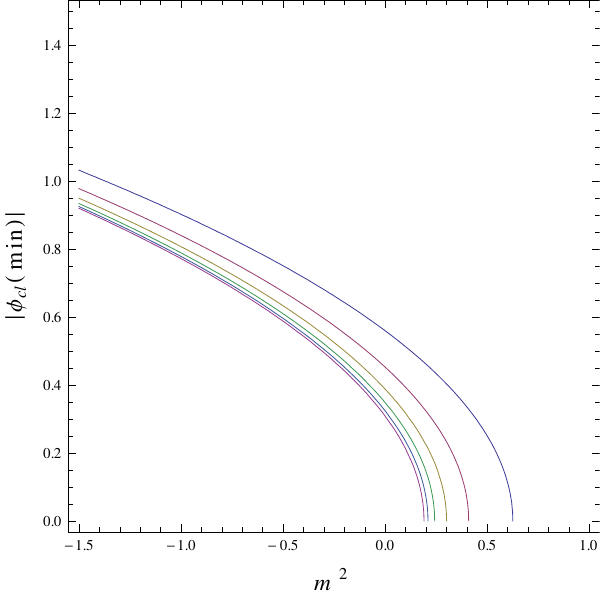}
\caption{$M^2=0$ plots for values of $\b$ in the range $[1,2]$ on the $m^2$-$|\phi^{}_{cl}(\mbox{min})|$ plane. }
\label{contourymin3}
\end{subfigure}
\caption{Bounded/unbounded effective potentials and $M^2=0$ plots for $AdS_3$ with negative renormalized volume.}
\label{ads3fig4}
\end{figure}

\begin{figure}[h]
\begin{subfigure}[b]{0.5\linewidth}
    \centering
\includegraphics[width= 11cm,angle=0]{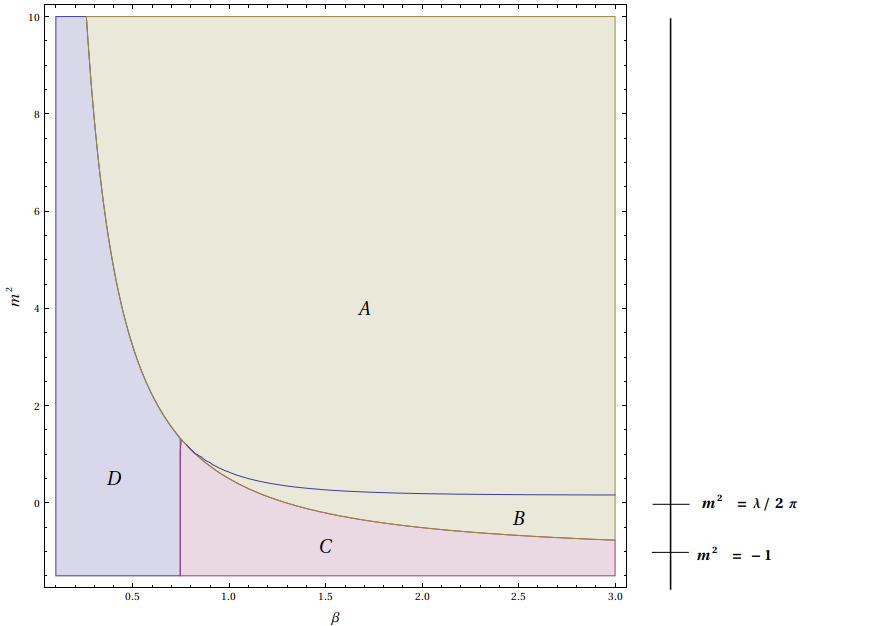}
\caption{Phases on the $\b-m^2$ plane.}
\label{ads3regions}
\end{subfigure}% \hspace{-2em}
\begin{subfigure}[b]{0.5\linewidth}
\centering
\includegraphics[width=8cm]{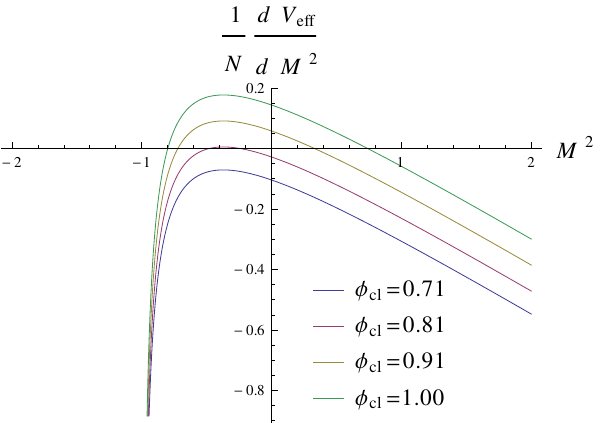}
\caption{Roots of the saddle point equation, $m^2=-0.8$, $\beta = 1$, $n=10$ and different values of $\phi^{}_{cl}$. }
\label{roots}
\end{subfigure}
\caption{Phases and roots of saddle point equation for $AdS_3$ with negative renormalized volume.}
\label{ads3fig5}
\end{figure}

We first consider the case when the regularized volume is negative (\ref{vol3}). The above equation can be solved numerically by restricting the sum to the first few values of $n$. In the following we have done the analysis for $n=10$ which gives quite good convergence. Relevant plots are shown in figures \ref{ads3fig4} and \ref{ads3fig5}.

There are two roots for $M^{2}$ from the saddle point equation (\ref{saddle3}), one moves towards $M^{2} = -1$ and the other away in the positive direction as $\phi^{}_{cl}$ is increased. This behaviour of the roots of the saddle point equation is depicted in Figure \ref{roots}.

It is observed from the expression for the effective potential, that corresponding to the root of $M^{2}$ from the saddle point equation that goes to  $-1$, the value of $V^{}_{eff}$ becomes unbounded below as $\phi^{}_{cl}$ is increased. On this physical ground we can discard the left-hand-side root where $M^{2}$ goes to $-1$. Effective potentials corresponding to these two roots are shown in Figure \ref{potential}.

\begin{figure}[H] 
  \begin{subfigure}[b]{0.5\linewidth}
    \centering
    \includegraphics[width=0.75\linewidth]{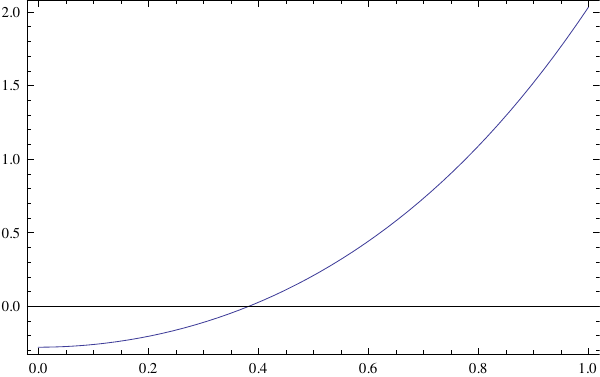} 
    \caption*{$A$: $m^2=4.0$,$\b=1.7$} 
    %\label{a} 
    \vspace{1ex}
  \end{subfigure}%
   \hspace{-2em}
  \begin{subfigure}[b]{0.5\linewidth}
    \centering
    \includegraphics[width=0.75\linewidth]{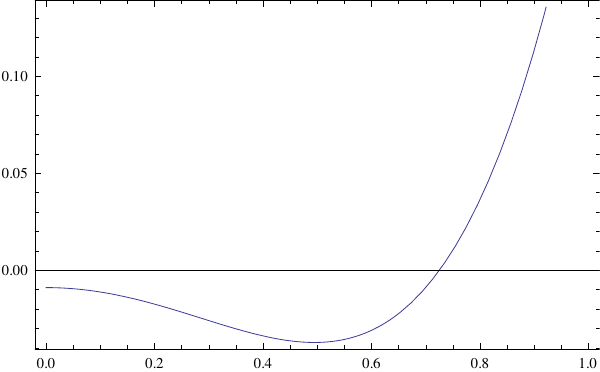} 
    \caption*{$B$: $m^2=-0.32$,$\b=2.5$} 
    %\label{b} 
    \vspace{1ex}
  \end{subfigure} 
  \begin{subfigure}[b]{0.5\linewidth}
    \centering
    \includegraphics[width=0.75\linewidth]{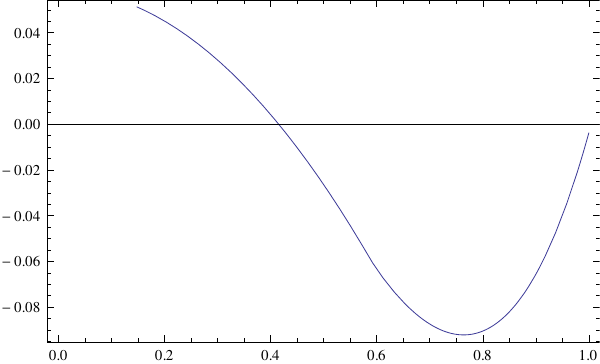} 
    \caption*{$C$: $m^2=-0.9$,$\b=1.5$} 
    %\label{c} 
  \end{subfigure}%%
  \begin{subfigure}[b]{0.5\linewidth}
    \centering
    \includegraphics[width=0.75\linewidth]{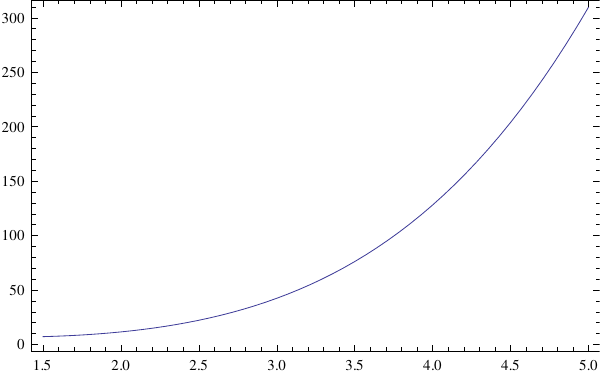} 
   \caption*{$D$: $m^2=0.5$,$\b=0.4$} 
    %\label{d} 
  \end{subfigure} 
  \caption{Representative plots of the potential corresponding to the regions in Figure \ref{ads3regions} for $\l=1$. }
  \label{fig5} 
\end{figure}

 We summarize the nature of the potentials in this theory for various values of the parameters. Figure \ref{ads3regions} shows the phases of the theory in the $\b-m^2$ plane. (i) $A$: Minimum only at $|\phi_{cl}|=0$ (ii) $B$: Maximum at $|\phi_{cl}|=0$ and a minimum at $|\phi_{cl}|\ne 0$. Figure \ref{contourymin3} shows the values of $|\phi_{cl}(\mbox{min})|$ versus $m^2$ for various $\beta$. Each contour satisfies $M^2=0$. The inner and outer contours are for $\b=2$ and $\b=1$ respectively. The region $B$ shrinks for small values of $\b$ (high temperature) and expands to a width given by $-1 < m^2 \le \l/(2\pi)$ at zero temperature. The upper limit is obtained from the saddle point equation (\ref{saddle3}) at zero temperature by putting $M^2=|\phi_{cl}|=0$. The the lower limit of $m^2$ can be obtained as follows. The RHS of the saddle point equation (\ref{saddle3}) at zero temperature is a monotonically decreasing function of $M^2$  which ends when $M^2=-(d/2)^2$. Real roots of  $M^2$ will exist as long as the value of the function at the endpoint is positive. Since the term containing $m^2$ in (\ref{saddle3}) is independent of $M^2$, the limiting value of $m^2$ can be obtained by putting $M^2=-(d/2)^2$ in the saddle point equation. Since $m^2 \ge -1$ is above the BF bound, both the symmetry breaking and the symmetry preserving phases coexist in this region. (iii) $C$: Only the $|\phi_{cl}|\ne 0$ minimum exists. The effective potential plots end when there exists no solution for the saddle point equation. This may be seen graphically in the Figure \ref{roots}. There exist solutions of the saddle point equation only beyond a certain value of $|\phi^{}_{cl}|$. (iv) $D$: There appears to be no ground state accessible by the present analysis. In this region too the there is no solution to the saddle point equation below a certain value of $|\phi_{cl}|$ satisfying the extremum condition $\pa V_{eff}/\pa |\phi _{cl}|=|\phi _{cl}|M^2=0$. Plots corresponding to various $m^2$ and $\b$ values are shown in Figure \ref{fig5}.
 
We now discuss some of the details of the numerics. The boundary separating the phases $A$ and $B$ is the contour on which the two extrema coincide, requiring $\pa^2 V_{eff}/\pa |\phi _{cl}|^2=M^2=0$. Numerically this is given by the contour $M^2=0$ root of the saddle point equation (\ref{saddle3}) for $|\phi_{cl}|=0$. The phase boundary between $B$ and $C$ and also $A$ and $D$ is the contour for $|\phi_{cl}|=0$ below which there is no real solution for $M^2$ from the saddle point equation. From the concave nature of the plot of the RHS of the saddle point equation Figure \ref{roots}, the boundary values of $(\b, m^2)$ corresponding to the limiting condition for existence of real $M^2$ solutions is obtained when the curve is tangential to the $M^2$ axis i.e.

\begin{figure}[h]
\begin{subfigure}[b]{0.5\linewidth}
    \centering
\includegraphics[width= 10cm,angle=0]{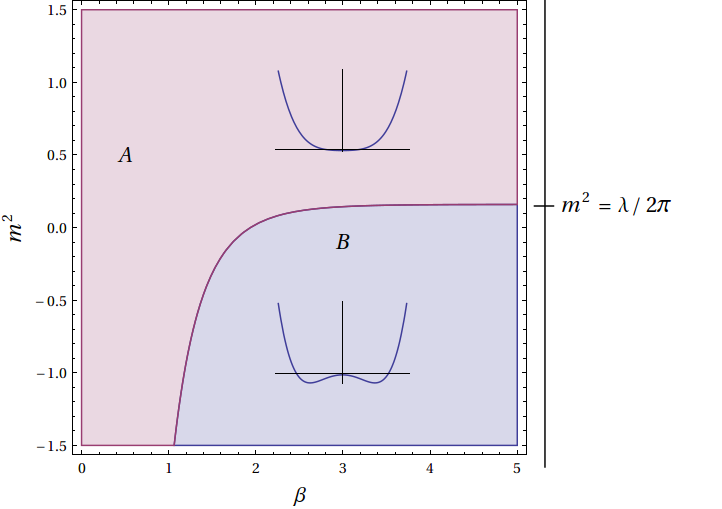}
\caption{Phases on the $\b-m^2$ plane.}
\label{ads3regionspos}
\end{subfigure}% \hspace{-2em}
\begin{subfigure}[b]{0.5\linewidth}
\centering
\includegraphics[width=7cm]{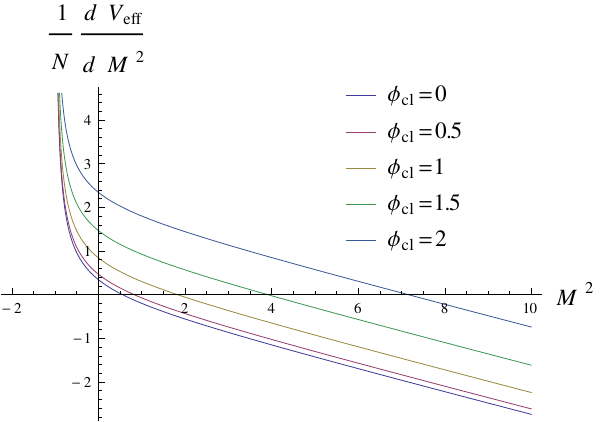}
\caption{Roots of the saddle point equation, $m^2=-0.5$, $\beta = 1$, $n=10$ and different values of $\phi^{}_{cl}$. }
\label{rootspos}
\end{subfigure}
\caption{Phases and roots of saddle point equation for $AdS_3$ with positive renormalized volume.}
\label{phaseads3pos}
\end{figure}

\beqa\label{saddle3d}
\frac{\pa}{\pa M^2}\left[\frac{m^{2}-M^{2}}{4\lambda}-\frac{\sqrt{1+M^{2}}}{8\pi}-\f{1}{\pi}\sum^{\infty}_{n=1}\frac{e^{-n\beta\left(1+\sqrt{1+M^2}\right)}}{|1-e^{2\pi i n \t}|^{2}\sqrt{1+M^{2}}}\right]=0
\eeqa

The boundary separating $C$ and $D$ is the contour on the left of which $M^2=0$ solution ceases to exist. From the saddle point equation 
(\ref{saddle3}) it can be seen that this condition depends only on $\b$. The boundary value of $\b$ can be obtained from (\ref{saddle3d}) by setting $M^2=0$. For $\th=0$, this limiting value is $\b_l=0.745$.

Let us now consider the case when the regularized volume is positive (\ref{dimregv}). Relevant plots are shown in figure \ref{phaseads3pos}. The phase plot is given in Figure
\ref{ads3regionspos}. Unlike the previous analysis where the renormalized volume was 
negative, here the saddle point equation has a solution for all values of $\phi_{cl}$, 
  $m^2$ and $\b$. This is because the RHS of the saddle point equation (\ref{saddle3}) as shown in Figure \ref{rootspos} is monotonically decreasing and the plot asymptotes to positive infinity for $M^2=-1$. The corresponding $C$ and $D$ regions of Figure \ref{ads3regions} are thus absent here. The phase plot is simpler and only has two regions corresponding to the sign of the second derivative of $V_{eff}$ with respect to  $\phi_{cl}$ (which is $M^2$) at $\phi_{cl}=0$. The phase boundary given by $M^2 = 0$ asymptotes to $m^2 = \lambda/(2\pi)$.  

\subsubsection{$AdS_2$}\label{ads2ln}

The zero temperature (Euclidean $AdS_2$) contribution to the trace can be obtained from (\ref{trzt}). Since the trace is divergent we expand about $d=1$ as in (\ref{expads2}). To renormalize the effective potential we include a mass counterterm so that

\begin{equation}\label{efflnct}
\f{V_{eff}(\phi_{cl}^i,\s_{cl})}{N}=-\frac{(M^{2}-m^2)^2}{8\l}+\frac{M^{2}}{2}(\phi^{i}_{cl})^{2}+M^2\f{\d m^2}{4\l}-\f{1}{{\cal V}_{d+1}}(\log Z^{(1)}+\log Z^{(1)}_{\t}).
\end{equation}

and define the renormalized mass as 

\beqa
\left.\frac{1}{N}\frac{\partial }{\partial M^{2}}V^{0}_{eff}(\phi_{cl}^i,\s_{cl})\right|_{M^2=\phi_{cl}^i=0}=\f{m^2}{4\l}.
\eeqa

where $V^{0}_{eff}$ in this equation is for zero temperature. This renormalization condition gives

\beqa
\f{\d m^2}{4\l}=-\f{{\m}^{-\e}}{2{\cal V}_{d+1}}\tr \left[\frac{1}{-\square_E}\right]=-\f{1}{4\pi\e}-\f{1}{8\pi}\left[\g +\log (4\pi) -\log({\m}^2)\right]+{\cal O}(\e)
\eeqa

\begin{figure}[H]
\begin{subfigure}[b]{0.5\linewidth}
    \centering
\includegraphics[width= 12cm,angle=0]{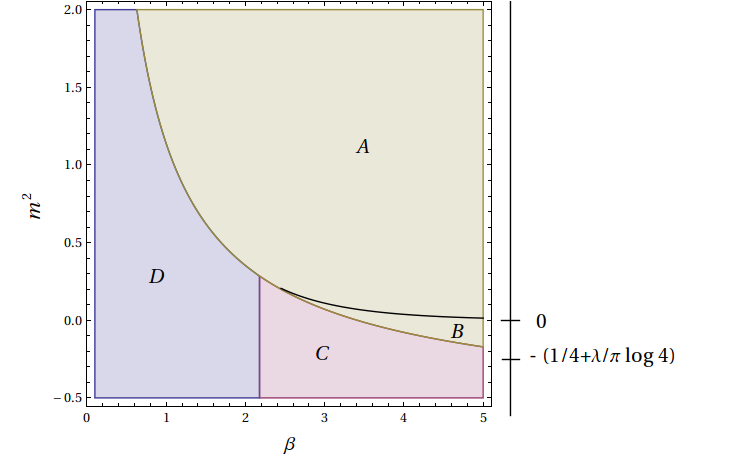}
\caption{Phases in $AdS_2$ on the $\b-m^2$ plane.}
\label{ads2regions}
\end{subfigure}% \hspace{-2em}
\begin{subfigure}[b]{0.5\linewidth}
\centering
\includegraphics[width=7cm]{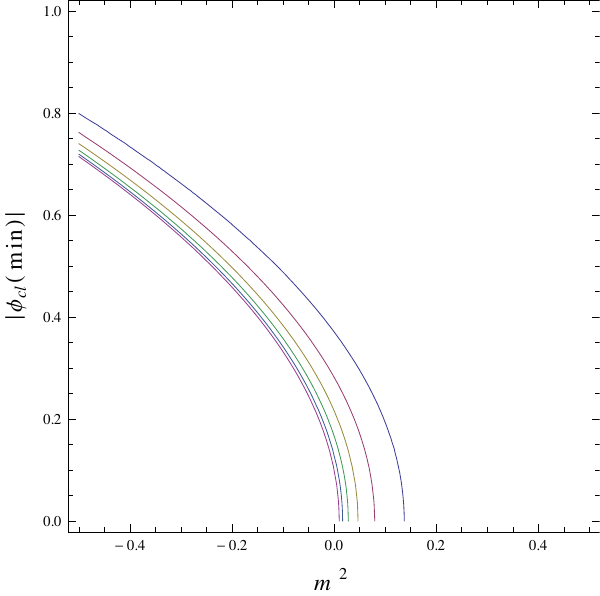}
\caption{$M^2=0$ plots for values of $\b$ in the range $[2.8,5.3]$ on the $m^2$-$|\phi^{}_{cl}(\mbox{min})|$ plane.  }
\label{roots2}
\end{subfigure}
\caption{Phases and  $M^2=0$ plots for $AdS_2$.}
\label{ads2fig8}
\end{figure}

\begin{figure}[H] 
  \begin{subfigure}[b]{0.5\linewidth}
    \centering
    \includegraphics[width=0.75\linewidth]{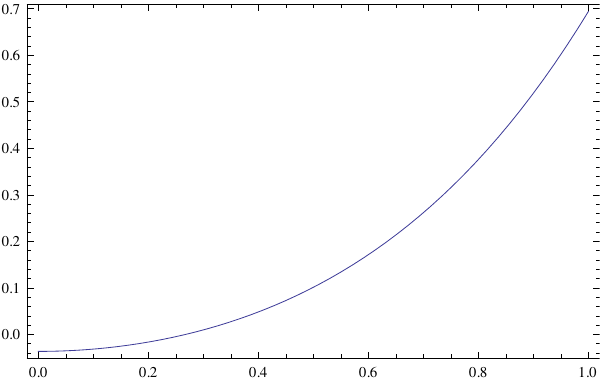} 
    \caption*{$A$: $m^2=1.1$, $\beta=3.5$} 
    %\label{a} 
    \vspace{1ex}
  \end{subfigure}%% 
  \begin{subfigure}[b]{0.5\linewidth}
    \centering
    \includegraphics[width=0.75\linewidth]{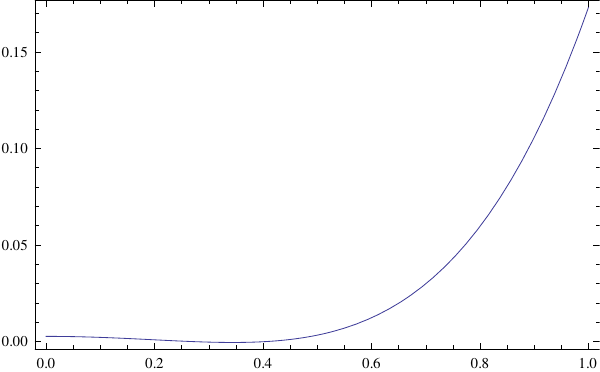} 
    \caption*{$B$: $m^2=-0.1$, $\beta=4.7$} 
    %\label{b} 
    \vspace{1ex}
  \end{subfigure} 
  \begin{subfigure}[b]{0.5\linewidth}
    \centering
    \includegraphics[width=0.75\linewidth]{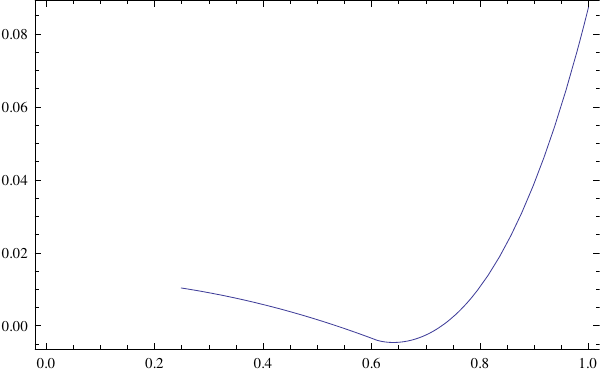} 
    \caption*{$C$: $m^2=-0.3$, $\beta=3.0$} 
    %\label{c} 
  \end{subfigure}%%
  \begin{subfigure}[b]{0.5\linewidth}
    \centering
    \includegraphics[width=0.75\linewidth]{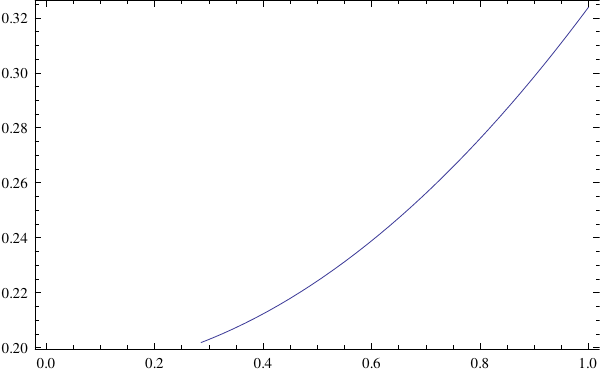} 
   \caption*{$D$: $m^2=-0.4$, $\beta=1.2$} 
    %\label{d} 
  \end{subfigure} 
  \caption{Representative plots of the potential corresponding to the regions in Figure \ref{ads2regions} for $\lambda=0.5$. }
  \label{fig7} 
\end{figure}

Inserting the finite temperature contribution from (\ref{zdt0}), ${\cal V}^{}_{2}=-\beta$ from (\ref{vol2}) and after removing an infinite constant, the effective potential at finite temperature becomes
\begin{eqnarray}
\frac{V^{}_{eff}}{N}=-\frac{(M^{2}-m^{2})^{2}}{8\lambda}+\frac{1}{2}M^{2}(\phi^{i}_{cl})^{2}-\frac{1}{4\pi}\int^{M^{2}}_{0} dM^2 \left[ \psi^{(0)}\left(\n+ \frac{1}{2}\right)+\g\right] \nonumber \\
+\f{1}{\b}\sum^{\infty}_{n=1}\frac{1}{n}\frac{e^{-n\beta\left(\frac{1}{2}+\sqrt{\frac{1}{4}+M^2}\right)}}{|1-e^{-n\beta}|}
\end{eqnarray}

and the saddle point equation is

\begin{eqnarray}\label{sads2}
0=\frac{1}{N}\frac{\partial V}{\partial M^{2}}=\frac{m^{2}-M^{2}}{4\lambda}+\f{(\phi^{i}_{cl})^{2}}{2}-\frac{1}{4\pi}\psi^{(0)} \left(\frac{1}{2}+\sqrt{\frac{1}{4}+M^{2}}\right)-\frac{\g}{4\pi}\nonumber\\ 
-\f{1}{2}\sum^{\infty}_{n=1}\frac{e^{-n\beta\left(\frac{1}{2}+\sqrt{\frac{1}{4}+M^2}\right)}}{|1-e^{-n\beta}|\sqrt{\frac{1}{4}+M^{2}}}
\end{eqnarray}

The relevant plots are shown in figure \ref{ads2fig8}. The phase plot for the theory on $AdS_2$, Figure \ref{ads2regions} is similar to that of $AdS_3$ with negative regularized volume (Figure \ref{ads3regions}). Here too there are four regions marked 
$A$-$D$ corresponding to the existence of the extrema. The phase boundaries are obtained numerically (for $n=10$) using similar observations as in section \ref{ads3n}. The coexistence region asymptotes to $-(1/4 +\l/\pi\log4) < m^2 \le 0 $ at zero temperature.

 The upper limit is determined from the saddle point equation (\ref{sads2}) by inserting $|\phi_{cl}|=M^2=0$ and the lower limit is obtained by requiring that real solution of $M^2$ exists for $|\phi_{cl}|=0$ in (\ref{sads2}). Contour plots for $M^2=0$ on the
$m^2$-$|\phi^{}_{cl}(\mbox{min})|$ plane for various values of $\b$ are shown in Figure \ref{roots2}.   
The limiting value $\b_l$ below which there is no solution to the saddle point equation corresponding to an extremum is $2.178$ ($\l=0.5$). The existence of the symmetry breaking phase in this theory  as noted in \cite{Inami:1985dj}, in two-dimensional $AdS$ space unlike that in flat-space is not a large $N$ phenomenon, but is due to the curvature of the background $AdS$ space. The long-range two-point correlator for the fields in the symmetry broken phase remain finite even for finite $N$. Plots of potentials corresponding to various regions $A-D$ are shown in Figure \ref{fig7}.

\subsubsection{$AdS_4$}

Expanding the expression for the trace at zero temperature as (\ref{zads4exp}) and adding counterterms corresponding to $m^2$ and $\l$, 

\begin{eqnarray}\label{efflnct4}
\f{V_{eff}(\phi_{cl}^i,\s_{cl})}{N} &=&-\frac{(M^{2}-m^2)^2}{8\l}+\frac{M^{2}}{2}(\phi^{i}_{cl})^{2}-\f{1}{{\cal V}_{d+1}}(\log Z^{(1)}+\log Z^{(1)}_{\b})\non
&+& M^2\f{\d m^2}{4\l}-M^4\d\left(\f{1}{8\l}\right).
\end{eqnarray}

With the renormalization conditions (at zero temperature),

\beqa
\left.\frac{1}{N}\frac{\partial }{\partial M^{2}}V^{0}_{eff}(\phi_{cl}^i,\s_{cl})\right|_{M^2=\phi_{cl}^i=0}=\f{m^2}{4\l}
\eeqa
\beqa
\left.\frac{1}{N}\frac{\partial^2 }{\partial (M^{2})^2}V^{0}_{eff}(\phi_{cl}^i,\s_{cl})\right|_{M^2=0}=-\f{1}{4\l}
\eeqa

we can compute the counterterms to be

\beqa
\f{\d m^2}{4\l}=-\f{{\m}^{-\e}}{2{\cal V}_{d+1}}\tr \left[\frac{1}{-\square_E}\right]=\f{1}{8\pi^2\e}+\f{1}{16\pi^2}\left[\g +\log (4\pi)-\frac{1}{2}-\log({\m}^2)\right]+{\cal O}(\e)
\eeqa

\beqa
\d\left(\f{1}{4\l}\right)&=&\f{{\m}^{-\e}}{2{\cal V}_{d+1}}\f{\pa}{\pa M^2}\tr \left[\frac{1}{-\square_E+M^2}\right]_{M^2=0}\non
&=&-\f{1}{16\pi^2\e}-\f{1}{32\pi^2}\left[\g +\log (4\pi)-\frac{1}{2}-\log({\m}^2)\right]\non
&+& \f{1}{48\pi^2}[\psi^{(1)}(1)+\psi^{(1)}(3)]+{\cal O}(\e)
\eeqa

We can now write down the renormalized effective potential at zero temperature,

\beqa\label{vads4}
\frac{V(M^{2},\phi^{i}_{cl})}{N}=-\frac{(M^{2}-m^{2})^{2}}{8\lambda}+\frac{1}{2}M^{2}(\phi^{i}_{cl})^{2}&+&\frac{1}{2}\int^{M^{2}}_{0} dM^2~ \tr \left[\frac{1}{-\square_E+M^2}\right]_{\mbox{ren}}\non
&-&\f{M^4}{96\pi^2}[\psi^{(1)}(1)+\psi^{(1)}(3)]
\eeqa

where,

\beqa
\tr \left[\frac{1}{-\square_E+M^2}\right]_{\mbox{ren}}=\frac{(2+M^2)}{16\pi^2}\left[\psi^{(0)}\left(\n-\f{1}{2}\right)+\psi^{(0)}\left(\n+\f{3}{2}\right)+2\g-\f{3}{2}\right].
\eeqa

The finite temperature effective potential can be written by including the partition function for thermal $AdS^{}_{4}$ from (\ref{zdt0}) with the regularized volume obtained from (\ref{vol4}) which is ${\cal V}^{}_{4}=2\pi\beta/3$.

The saddle point equation is
\begin{eqnarray}\label{saddle4}
0=\frac{1}{N}\frac{\partial V}{\partial M^{2}}&=&\frac{m^{2}-M^{2}}{4\lambda}+\frac{(\phi^{i}_{cl})^{2}}{2}+ 
\frac{(2+M^2)}{32\pi^2}\left[\psi^{(0)}\left(\n-\f{1}{2}\right)+\psi^{(0)}\left(\n+\f{3}{2}\right)+2\g-\f{3}{2}\right]\non
&-&\f{M^2}{48\pi^2}[\psi^{(1)}(1)+\psi^{(1)}(3)]+\frac{3}{4\pi}\sum^{\infty}_{n=1}\frac{e^{-n\beta(\frac{3}{2}+\sqrt{\frac{9}{4}+M^2})}}{|1-e^{-n\beta}|^{3}\sqrt{\frac{9}{4}+M^{2}}}
\end{eqnarray}

\begin{figure}[H]
\begin{subfigure}[b]{0.5\linewidth}
    \centering
\includegraphics[width=8cm,angle=0]{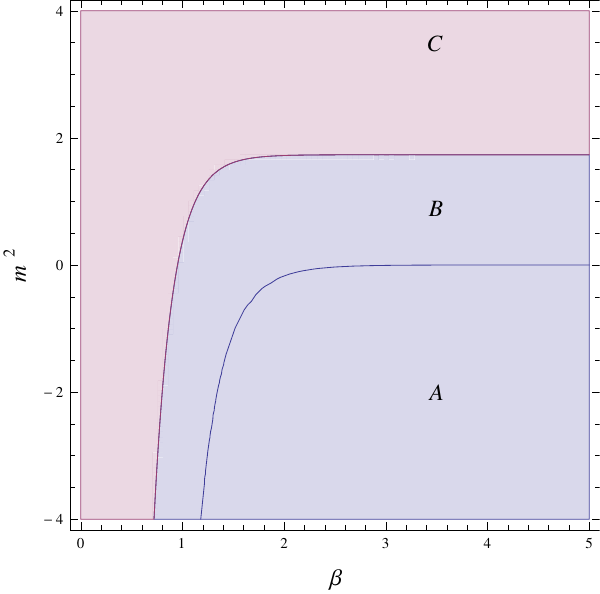}
\caption{Phases on $AdS_4$.}
\label{ads4regions}
\end{subfigure}% \hspace{-2em}
\begin{subfigure}[b]{0.5\linewidth}
\centering
\includegraphics[width=8cm]{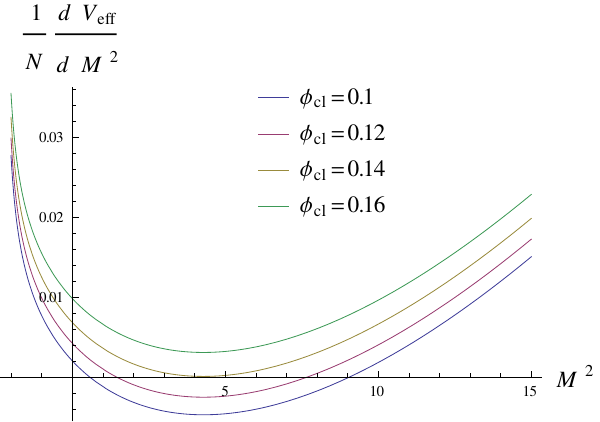}
\caption{Roots of the saddle point equation, $m^2=-1$, $\beta = 2$, $n=10$ and different values of $\phi^{}_{cl}$. }
\label{roots4}
\end{subfigure}
\caption{Phases and roots of saddle point equation for $AdS_4$.}
\label{ads4fig10}
\end{figure}

The zero temperature contribution to the RHS of the saddle point equation (\ref{saddle4}) is not a monotonic function but is convex unlike that of the previous $AdS_2$ and $AdS_3$ cases. The relevant plots are shown in figure \ref{ads4fig10}. Looking at the large $M$ behavior of the RHS, we note that the curve starts rising when $\log(M) \sim (8\pi^2)/\l$. This is why we need to choose a large value of $\l$ for the numerics which we fix to be $\l=70$. 
The convex nature of the 
RHS of saddle point equation (\ref{saddle4}) persists for finite temperature (see Figure \ref{roots4}). Further, since the coefficient of the term containing $|\phi_{cl}|$ is independent of $M$, there is no solution to the saddle point equation beyond a certain value of $|\phi_{cl}|$. The potential plots end beyond this value. This is unlike the previous concave nature where the potential plots end at lower values of $|\phi_{cl}|$. There is thus no physical ground on which we may concentrate on one of the roots.

We thus summarize the nature of the potentials in this theory corresponding to both the roots of the saddle point equation for various values of the parameters.

\noindent
(i) The phases are shown in Figure \ref{ads4regions}. There is no solution to the saddle point equation in region $C$ and the two roots
as shown in Figure \ref{roots4} coincide on the boundary separating regions $B$ and $C$.

\noindent
(ii) The second derivative of the potential (\ref{vads4}) with respect to $|\phi_{cl}|$ is $M^2$. Thus the sign of the root of the saddle point equation determines the nature of extremum (maximum or minimum). We find that for $|\phi_{cl}|=0$ the right root (as in Figure \ref{roots4}) is always positive and the corresponding potential is stable at $|\phi_{cl}|=0$ both in regions $A$ and $B$.

\noindent
 (iii) The left root of the saddle point equation gives a maximum at $|\phi_{cl}|=0$ and a minimum $|\phi_{cl}|\ne 0$ in region $A$. The two extrema coincide at $|\phi_{cl}|=0$ on the boundary separating $A$ and $B$.

 For $M^2=|\phi_{cl}|=0$, i.e, on boundary separating the regions $A$ and $B$, the $m^2$ value approaches zero for large $\b$ as in $AdS_2$. The corresponding value on the boundary separating regions $A$ and $C$ at zero temperature is obtained from the condition that the two roots of the saddle point equation coincide. For $\l=70$ this gives $m^2\sim 1.734$.

\section{Discussion}\label{summary}

In this paper we have given a derivation for the one-loop partition function $Z^{(1)}$ for scalars using the eigenfunctions of the Laplacian operator in Euclidean $AdS$. We have shown that our computation involves the method of images applied to the Green's function and generalizes to $AdS$ spaces with arbitrary dimensions thus reproducing results derived using other techniques \cite{Giombi:2008vd}-\cite{Gopakumar:2011qs},\cite{Keeler:2014hba},\cite{Kraus:2020nga}. For Euclidean $AdS$ we have also demonstrated the equality of the various integral representations of the trace (\ref{trace1}).  

Equipped with the above leading order results we studied the phases of scalar theories on thermal $AdS^{}_{d+1}$. Though inclusion of 
the angular potentials $\theta$ would result in a richer phase diagram, as an initial step we have set the angular potentials 
$\theta=0$. After an analysis of the theory with a single scalar field we studied the phases of $O(N)$ vector model for finite $N$ as well as for the large $N$ limit. We identified regions in the $\beta-m^2$ parameter space that correspond to the symmetry preserving and symmetry breaking phases for $d=1,2,3$, performing the analysis numerically. The zero temperature effective potentials are UV divergent and were renormalized using the standard procedure of dimensional regularization and inclusion of counterterms. We also used the regularized volume of thermal $AdS^{}_{d+1}$ to proceed with the analysis. 

One of the constraints in the determination of the phases in the large $N$, $O(N)$ vector model is the existence of the solution corresponding to the saddle point equation. The zero temperature contribution to the saddle point equation for $AdS_2$ and $AdS_3$ as a function of the effective mass-squared $M^2$ qualitatively differs from that of $AdS_4$. Along with this the sign of the regularized volume of thermal $AdS^{}_{d+1}$ leads to different qualitative nature of the phase diagrams. 
The plots of RHS of the saddle point equations in $AdS^{}_{2}$ and $AdS^{}_{3}$ with negative regularized volume being concave (Figure \ref{roots}) results in the potential plots to end at lower values of $|\phi^{}_{cl}|$. For $AdS_3$ with positive regularized volume, the saddle point equation has solution for all values of $m^2$, $\b$ and $\phi_{cl}$ (Figure \ref{rootspos}). For $AdS^{}_{4}$, where the plot for the RHS of the saddle point equation is convex, the potential plots end beyond a certain value of $|\phi^{}_{cl}|$ (Figure \ref{roots4}). As was shown for zero temperature in \cite{Inami:1985dj}, we are able to confirm that for a finite temperature theory in $AdS$ there occurs a symmetry breaking phase in two dimensions, which is in contrast to the flat space where the Coleman-Mermin-Wagner theorem prohibits continuous symmetry breaking \cite{Mermin:1966fe,Coleman:1973ci}.
 
It was observed that unlike the flat space, there exists a region in $AdS$ space where both the symmetry breaking and symmetry preserving phases coexist. With the present analysis, it has not been possible to study the nature of transition between these two phases as a function of temperature.  Further in our analysis, we found a region where there exists no solution corresponding to a minimum of the potential. The perturbative analysis carried out here could well be valid for small enough temperatures and masses. For a more realistic setup one should supplement the phase diagrams by including asymptotically $AdS$ black-holes for high temperatures. We would like to address these in future. 
 
The study of phases of the large $N$ model carried out here may be extended in various directions. We would like to understand the implications of the present study on the dual boundary theory. The analysis of correlation functions at finite temperatures is expected to shed more light on this aspect. It would also be interesting to consider the theories including those of fermions on other thermal spaces with maximal symmetry.       
\\
 
\vspace{2.5mm}
\noindent
{\bf Acknowledgements :}
\\ We are grateful to Balachandran Sathiapalan for useful discussions. We also thank an anonymous referee of the paper
for valuable comments and suggestions. Astha Kakkar acknowledges the support of Department of Science and Technology (DST), 
Ministry of Science and Technology, Government of India, for the DST INSPIRE Fellowship with the INSPIRE
Fellowship Registration Number: IF180721. S.S. thanks the University Grants Commission (UGC), and DST, New
Delhi, India, for providing special assistance and infrastructural support to the Department of Physics,
Vidyasagar University, through the SAP and FIST program respectively.

\appendix

\section{Performing integrals in various orders}\label{var}

\subsection*{Integral (\ref{trace1})}

\begin{eqnarray}
\frac{1}{L^{2}}\mbox{tr} \left(\frac{1}{-\square^{}_{E}+V''(\phi^{}_{cl})}\right)=\frac{1}{L^{d+1}}\int d^{d+1}x \sqrt{g} \int_0^{\infty} \frac{d\lambda \mu(\lambda)}{\lambda^{2}+\n^{2}} \int \frac{d^{d}(ky)}{(2\pi)^{d}} [K^{}_{i\lambda}(ky)]^{2}
\end{eqnarray}

Using the expression for the area of $d-1$ dimensional unit sphere

\begin{equation}
\label{em12}
\int \frac{d\Omega^{}_{d}}{(2\pi)^{d}}=\frac{2\pi^{d/2}}{\Gamma(d/2)}\frac{1}{(2\pi)^{d}},
\end{equation}
integrating over $k$ and plugging in the normalization measure then gives 

\begin{equation}
\frac{1}{L^{2}}\mbox{tr} \left(\frac{1}{-\square^{}_{E}+V''(\phi^{}_{cl})}\right)=\frac{{\cal V}^{}_{d+1}}{L^{d+1}2^{d+1} \pi^{\frac{d+1}{2}} \Gamma(\frac{d}{2}+\frac{1}{2})}\int_{-\infty}^{\infty}\frac{d\lambda}{\lambda^{2}+\n^{2}}  \frac{\Gamma(\frac{d}{2} \pm i\lambda)}{\Gamma(\pm i\lambda)}
\end{equation}
where $\Gamma(\pm a)=\Gamma(a)\Gamma(-a)$.

Next the $\lambda$ integral needs to be performed. The poles for $\lambda$ occur when
\begin{equation}
\lambda^{2}+\n^{2}=0
\end{equation}
Therefore $\lambda=\pm i\n$. We close the contour in the upper half so that $\lambda=+ i\n$ and the residue from this pole is 
\begin{eqnarray}
\mbox{Residue}
=-\Gamma\left(\frac{d}{2}\pm \n\right) \sin(\pi \n)
\end{eqnarray}
The Gamma functions also give poles when
\begin{equation}
\frac{d}{2}\pm i\lambda=-n
\end{equation}
for $n=0,1,2, \cdots$.
However only $\Gamma(\frac{d}{2}+i\lambda)$ has poles in the upper half. Therefore
\begin{equation}
(2 \pi i)\mbox{Residue}=(2 \pi i)\left[\sum^{\infty}_{n=0} \frac{1}{\n^{2}-\left(\frac{d}{2}+n\right)^{2}}\frac{(-1)^{n}}{n!} \Gamma(d+n)i \left(n+\frac{d}{2}\right) \frac{\sin(\pi(n+\frac{d}{2}))}{\pi}\right].\end{equation}
This simplifies to
\begin{equation}
\frac{(-2) \pi^{2} \cos(\pi \n) \tan(\frac{d\pi}{2})}{(\cos(\pi d)-\cos(2\pi \n))\Gamma((1-(\frac{d}{2}+\n))\Gamma(1-(\frac{d}{2}-\n))}.
\end{equation}
After adding the two contributions we get,
\begin{equation}
\label{em1}
\frac{1}{L^{2}}\mbox{tr} \left(\frac{1}{-\square^{}_{E}+V''(\phi^{}_{cl})}\right)=\frac{{\cal V}^{}_{d+1}}{4(\pi)^{\frac{d+2}{2}}} \frac{\Gamma(\frac{d}{2})}{\Gamma(d)}\frac{1}{L^{d+1}}\frac{\Gamma(\frac{d}{2}\pm \n) \sin(\pi(\frac{d}{2}-\n))}{\cos(\frac{\pi d}{2})}.
\end{equation}

\subsection*{Integral (\ref{order3})}

Performing the $k$ integral and then the $s$ integral gives

\begin{equation}
\frac{1}{L^{2}}\mbox{tr} \left(\frac{1}{-\square^{}_{E}+V''(\phi^{}_{cl})}\right)={\cal V}_{d+1}\int\frac{d\Omega^{}_{d}}{(2\pi)^{d}}\frac{1}{4}\Gamma^{2}(d/2) 2^{d/2}\frac{1}{(2\pi i)}\int^{\infty+i\pi}_{\infty -i\pi} e^{-\n t}[1-\cosh \ t]^{-d/2}dt
\end{equation}
Now consider the integral
\begin{equation}
\frac{1}{(2\pi i)}\int^{\infty+i\pi}_{\infty -i\pi} e^{-\n t}[1-\cosh \ t]^{-d/2}dt
\end{equation}
over the contour as shown in Figure {\ref{f1}.
\begin{figure}[H]
\centering
\includegraphics[width=8cm,height=6cm]{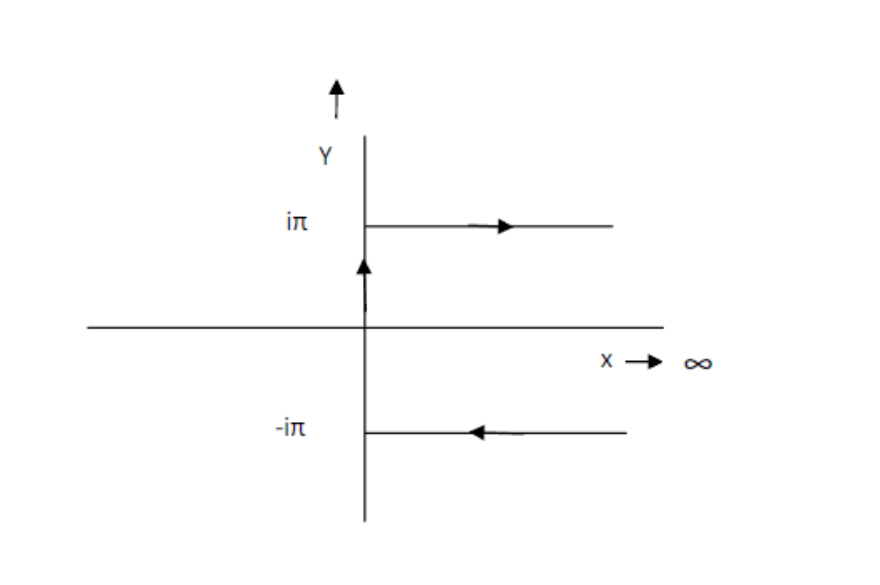}
\caption{The contour for integration over $t$.}
\label{f1}
\end{figure}
The three contributions to the integral are
\begin{eqnarray}
\frac{1}{(2\pi i)}\left(\int^{\infty}_{0} e^{-\n(i\pi+x)}[1-\cosh \ (i\pi+x)]^{-d/2}dx + \nonumber \right. \\
\int^{0}_{\infty} e^{-\n(-i\pi+x)}[1+\cosh(-i\pi+x)]^{-d/2}dx + \nonumber\\
\left. i\int^{\pi}_{-\pi} e^{-i\n x}[1-\cos \ x]^{-d/2}dx \right)
\end{eqnarray}
These integrals can be computed and plugged back into the main expression to give the trace as
\begin{eqnarray}
\label{e8a}
\frac{1}{L^{2}}\mbox{tr} \left(\frac{1}{-\square^{}_{E}+V''(\phi^{}_{cl})}\right)=\frac{{\cal V}^{}_{d+1}}{L^{d+1}}\frac{\Gamma(d/2)}{2\pi^{d/2}}\left[\frac{\sec(\frac{\pi d}{2})\Gamma(\frac{d}{2}-\n)}{2\Gamma(d)\Gamma(1-\frac{d}{2}-\n)}-\right.\nonumber \\
\left.\frac{\sin(\pi \n)}{\pi}\left(\Gamma\left(\n+\frac{d}{2}\right) \ ^{}_{2}F^{}_{1R}\left( \n+\frac{d}{2},d,1+\n+\frac{d}{2};-1\right)+\right.\right.\nonumber\\
\left.\left. \Gamma\left(-\n+\frac{d}{2}\right) \ ^{}_{2}F^{}_{1R}\left(d,-\n+\frac{d}{2},1-\n+\frac{d}{2};-1\right)\right)\right]
\end{eqnarray}
where 
\begin{equation}\label{hyr}
^{}_{2}F^{}_{1R}(a,b,c;z)=\frac{^{}_{2}F^{}_{1}(a,b,c;z)}{\Gamma(c)}
\end{equation}
is the regularized hypergeometric function.

\subsection*{Integral (\ref{order2})}

The integral over $k$ gives
\begin{equation}
\label{e7}
\int dk k^{d-1}\exp[-sk^{2}]=\frac{1}{2}s^{-d/2}\Gamma\left(\frac{d}{2}\right)
\end{equation}
Then finally the integral over $s$ and the use of the relation in equation (\ref{em12}) gives
\begin{multline}
\label{e8}
\frac{1}{L^{2}}\mbox{tr} \left(\frac{1}{-\square^{}_{E}+V''(\phi^{}_{cl})}\right)=\frac{{\cal V}^{}_{d+1}}{L^{d+1}} \frac{\Gamma(1/2)}{4\pi^{d/2}}\frac{\Gamma(\frac{1}{2}-\frac{d}{2})}{\Gamma(\frac{1}{2}-\frac{d}{4}+\frac{\n}{2})\Gamma(1-\frac{d}{4}+\frac{\n}{2})}\times  \\
\left(\frac{\cos(\frac{\pi}{4}(d+2\n))\Gamma(\frac{d+2\n}{4})}{\Gamma(\frac{1}{2}-\frac{d}{4}-\frac{\n}{2})} 
+\frac{\sin(\frac{\pi}{4}(d+2\n))\Gamma(\frac{2+d+2\n}{4})}{\Gamma(1-\frac{d}{4}-\frac{\n}{2})}  \right)
\end{multline}

We may check the equality of the four expressions for the trace by using different properties of the special functions involved as shown below.
 \\
 
\subsection*{Equality of the expressions (\ref{trzt}), (\ref{em1}), (\ref{e8a}) and (\ref{e8})}

\subsubsection*{ Expression (\ref{e8})}

We use the following relations for Gamma functions
\begin{equation}
\label{e9}
\Gamma(z)\Gamma(z+\frac{1}{2})=2^{1-2z}\sqrt{\pi}\Gamma(2z)
\end{equation}
and
\begin{equation}
\label{e10}
\Gamma(1-z)\Gamma(z)=\frac{\pi}{\sin(\pi z)}
\end{equation}
Thus the term in the brackets in equation(\ref{e8}) can be written as
\begin{multline}
\label{e11}
\frac{\cos\left(\frac{\pi}{4}\left(d+2\n\right)\right)\Gamma\left(\frac{1}{4}\left(d+2\n\right)\right)\Gamma\left(1-\frac{d}{4}-\frac{\n}{2}\right)}{\Gamma\left(1-\frac{d}{4}-\frac{\n}{2}\right)\Gamma\left(\frac{1}{2}-\frac{d}{4}-\frac{\n}{2}\right)}+ \\
 \frac{\sin\left(\frac{\pi}{4}\left(d+2\n\right)\right)\Gamma\left(\frac{1}{2}+\frac{d}{4}+\frac{\n}{2}\right)\Gamma\left(\frac{1}{2}-\frac{d}{4}-\frac{\n}{2}\right)}{\Gamma\left(1-\frac{d}{4}-\frac{\n}{2}\right)\Gamma\left(\frac{1}{2}-\frac{d}{4}-\frac{\n}{2}\right)}
\end{multline}
and using the relation in equation (\ref{e10}) we may write the second term in the numerator of equation(\ref{e11}) as
\begin{equation}
\sin\left(\frac{\pi}{4}\left(d+2\n\right)\right)\Gamma\left(\frac{1}{4}\left(2+d+2\n\right)\right)\Gamma\left(\frac{1}{2}-\frac{d}{4}-\frac{\n}{2}\right)=\frac{\sin\left(\frac{\pi}{4}\left(d+2\n\right)\right)\pi}{\cos\left(\frac{\pi}{4}\left(d+2\n\right)\right)}
\end{equation}
Thus numerator of equation(\ref{e11}) becomes
\begin{equation}
\frac{\pi\left(\cos^{2}\left(\frac{\pi}{4}\left(d+2\n\right)\right)+\sin^{2}\left(\frac{\pi}{4}\left(d+2\n\right)\right)\right)}{\cos\left(\frac{\pi}{4}\left(d+2\n\right)\right)\sin\left(\frac{\pi}{4}\left(d+2\n\right)\right)}
\end{equation}
Therefore the expression for trace becomes
\begin{eqnarray}
\f{1}{L^2}\mbox{tr} \left(\frac{1}{-\square^{}_{E}+V''(\phi^{}_{cl})}\right)=\f{2{\cal V}_{d+1}\pi^{\frac{1-d}{2}}}{L^{d+1}}\f{\Gamma\left(\frac{d}{2}+\n\right)\Gamma\left(1-\frac{d}{2}-\n\right)\Gamma\left(\frac{1}{2}-\frac{d}{2}\right)}{\Gamma\left(1-\frac{d}{4}-\frac{\n}{2}\right)\Gamma\left(\frac{1}{2}-\frac{d}{4}-\frac{\n}{2}\right)} \times \non
\frac{1}{4\Gamma\left(1-\frac{d}{4}+\frac{\n}{2}\right)\Gamma\left(\frac{1}{2}-\frac{d}{4}+\frac{\n}{2}\right)}
\end{eqnarray}
which on further simplification using property in equation(\ref{e9}) gives
\begin{equation}\label{a127}
\f{1}{L^2}\mbox{tr} \left(\frac{1}{-\square^{}_{E}+V''(\phi^{}_{cl})}\right)=\frac{{\cal V}^{}_{d+1}}{(4\pi)^{(d+1)/2}L^{d+1}}\frac{\Gamma\left(\frac{1}{2}-\frac{d}{2}\right)\Gamma\left(\frac{d}{2}+\n\right)}{\Gamma\left(1-\frac{d}{2}+\n\right)}
\end{equation}
It thus exactly matches with the expression in equation (\ref{trzt}).

\subsubsection*{Expression (\ref{em1})}

We may write relations for Gamma functions as
\begin{equation}
\frac{\Gamma\left(\frac{d}{2}\right)}{\Gamma(d)}=\frac{2^{1-d}\sqrt{\pi}}{\Gamma\left(\frac{1}{2}+\frac{d}{2}\right)}
\end{equation}
and
\begin{eqnarray}
\sin\left(\frac{\pi}{2}\left(1+d\right)\right)=\frac{\pi}{\Gamma\left(\frac{1}{2}+\frac{d}{2}\right)\Gamma\left(\frac{1}{2}-\frac{d}{2}\right)} 
= \cos\left(\frac{\pi d}{2}\right)
\end{eqnarray}
Substituting these values in equation(\ref{em1}) and simplifying we obtain (\ref{a127}) which again matches with the expression in equation (\ref{trzt}).

\subsubsection*{Expression (\ref{e8a})}
Using the relation (\ref{hyr}) for the regularized Hypergeometric function, the terms involving the hypergeometric functions in the numerator in equation (\ref{e8a}) can be simplified as
\begin{multline}
\label{e12}
\left(\frac{d}{2}-\n\right) \ ^{}_{2}F^{}_{1}\left(\n+\frac{d}{2},d,1+\n+\frac{d}{2},-1\right)+\\\left(\frac{d}{2}+\n\right) \ ^{}_{2}F^{}_{1}\left(d,\frac{1}{2}(-2\n+d),\frac{1}{2}(2-2\n+d),-1\right)
\end{multline}
Now let
\begin{equation}
a=\frac{d}{2}+\n~~~~\mbox{and}~~~~b=\frac{d}{2}-\n
\end{equation}

The permutation symmetry for hypergeometric functions states
\begin{equation}
\label{e13}
^{}_{2}F^{}_{1}\left(a,b,c,z\right)=\ ^{}_{2}F^{}_{1}\left(b,a,c,z\right)
\end{equation}
We further have the relation
\begin{equation}
\label{e14}
b\ ^{}_{2}F^{}_{1}\left(a,a+b,a+1,-1\right)+a\ ^{}_{2}F^{}_{1}\left(b,a+b,b+1,-1\right)=\frac{\Gamma(a+1)\Gamma(b+1)}{\Gamma(a+b)}
\end{equation}
Applying this symmetry property (\ref{e13}) and the relation in equation(\ref{e14}) to equation (\ref{e12}) we get
\begin{eqnarray}
&&\left(\frac{d}{2}-\n\right) \ ^{}_{2}F^{}_{1}\left(\n+\frac{d}{2},d,1+\n+\frac{d}{2},-1\right)+ 
\left(\frac{d}{2}+\n\right) \ ^{}_{2}F^{}_{1}\left(d,\frac{1}{2}(-2\n+d),\frac{1}{2}(2-2\n+d),-1\right) \nonumber \\ 
&=&\frac{\Gamma\left(\frac{d}{2}+\n+1\right)\Gamma\left(\frac{d}{2}-\n+1\right)}{\Gamma\left(d\right)}\end{eqnarray}
Therefore the complete term involving hypergeometric functions becomes
\begin{equation}
\frac{\Gamma\left(\frac{d}{2}+\n\right)\Gamma\left(\frac{d}{2}-\n\right)}{\Gamma\left(d\right)}
\end{equation}
Thus the terms in the bracket in expression(\ref{e8a}) become
\begin{equation}
\frac{\sec\left(\frac{\pi d}{2}\right)\Gamma\left(\frac{d}{2}-\n\right)}{2\Gamma(d)\Gamma\left(1-\frac{d}{2}-\n\right)} -\frac{\sin(\pi \n)}{\pi}\frac{\Gamma\left(\frac{d}{2}+\n\right)\Gamma\left(\frac{d}{2}-\n\right)}{\Gamma\left(d\right)}
\end{equation}

Using the property in equation(\ref{e10}) we get
\begin{equation}
\frac{\Gamma\left(\frac{d}{2}-\n\right)}{\Gamma(d)}\left(\frac{\pi-2\sin(\pi \n)\cos\left(\pi \frac{d}{2}\right)\frac{\pi}{\sin\left(\pi\left(\n+\frac{d}{2}\right)\right)}}{2\pi \cos\left(\pi\frac{d}{2}\right)\Gamma\left(1-\n-\frac{d}{2} \right)}\right)
\end{equation}

which further simplifies to

\begin{equation}
\frac{\Gamma\left(\frac{d}{2}-\n\right)}{2\Gamma(d)\cos\left(\pi\frac{d}{2}\right)\Gamma\left(1-\n-\frac{d}{2} \right)}\times\frac{\sin\left(\pi\left(\frac{d}{2}-\n\right)\right)}{\sin\left(\pi\left(\n+\frac{d}{2}\right)\right)}
\end{equation}
Simplifying further we get,
\begin{equation}
\f{1}{L^2}\mbox{tr} \left(\frac{1}{-\square^{}_{E}+V''(\phi^{}_{cl})}\right)=\frac{{\cal V}^{}_{d+1}}{L^{d+1}}\f{\Gamma\left(\frac{d}{2}-\n\right)\Gamma\left(\frac{d}{2}+\n\right)\sin\left(\pi\left(\frac{d}{2}-\n\right)\right)\Gamma\left(\frac{d}{2}\right)}{4(\pi)^{\frac{d}{2}+1}\Gamma\left(d\right)\cos\left(\frac{\pi d}{2}\right)}
\end{equation}
which is exactly the same as expression (\ref{em1})

\section{Delta function identities}\label{del}

To simplify the delta functions we use the following identity:

\beqa
\d^n(\vec{f}(\vec{x}))=\sum\f{\d^n(\vec{x}-\vec{x}_0)}{\left|\f{\pa(f_1,...,f_n)}{\pa(x_1,...,x_n)}\right|}
\eeqa 

where the sum runs over all the roots of the equations $\vec{f}(\vec{x})=0$.

Consider first the identity, equation (\ref{di1}): 

\beqa
\d^2(\g^n\vec{k}-\g^{n^{\prime}}\vec{k}^{\prime})&=&\d\left(e^{-n\beta}(k_1 \cos n\theta+k_2\sin n\theta)-e^{-n^{\prime}\beta}(k_1^{\prime} \cos n^{\prime}\theta+k^{\prime}_2\sin n^{\prime}\theta)\right)\non
&\times&\d\left(e^{-n\beta}(k_2 \cos n\theta-k_1\sin n\theta)-e^{-n^{\prime}\beta}(k^{\prime}_2 \cos n^{\prime}\theta-k^{\prime}_1\sin n^{\prime}\theta)\right)
\eeqa

\beqa
\left|\f{\pa(f_1,f_2)}{\pa(k^{\prime}_1,k^{\prime}_2)}\right|=\left|\begin{array}{cc}
-e^{-n^{\prime}\beta}\cos n^{\prime}\theta &-e^{-n^{\prime}\b}\sin n^{\prime}\theta\\
e^{-n^{\prime}\b}\sin n^{\prime}\theta&-e^{-n^{\prime}\beta}\cos n^{\prime}\theta\end{array}\right|=e^{-2n^{\prime}\b}
\eeqa

With the solutions $\vec{k}^{\prime}=\g^{n-n^{\prime}}\vec{k}$, we have the identity

\beqa\label{di1a}
\d^2(\g^n \vec{k}-\g^{n^{\prime}} \vec{k}^{\prime})=e^{2n^{\prime}\b}\d^2(\g^{(n-n^{\prime})} \vec{k}-\vec{k}^{\prime})
\eeqa

The other identity (\ref{di2}) is for $\vec{k}^{\prime}=\vec{k}$. The RHS of 
(\ref{di1a}) is 

\beqa
\d^2(\g^{(n-n^{\prime})} \vec{k}-\vec{k})&=&
\d\left(e^{-(n-n^{\prime})\beta}(k_1 \cos (n-n^{\prime})\theta+k_2\sin (n-n^{\prime})\theta)-k_1\right)\non
&\times&\d\left(e^{-(n-n^{\prime})\beta}(k_2 \cos (n-n^{\prime})\theta-k_1\sin (n-n^{\prime})\theta)-k_2\right)
\eeqa

\beqa
\left|\f{\pa(f_1,f_2)}{\pa(k_1,k_2)}\right|&=&\left|\begin{array}{cc}
e^{-(n-n^{\prime})\beta}\cos (n-n^{\prime})\theta-1 &e^{-(n-n^{\prime})\beta}\sin (n-n^{\prime})\theta\\
-e^{-(n-n^{\prime})\beta}\sin (n-n^{\prime})\theta&e^{-(n-n^{\prime})\beta}\cos (n-n^{\prime})\theta-1\end{array}\right|\non
&=&\left|1-2e^{-(n-n^{\prime})\b}\cos (n-n^{\prime})\theta+e^{-2(n-n^{\prime})\b}\right|\non
&=& \left|1-e^{2\pi i(n-n^{\prime})\tau}\right|^2
\eeqa

Since the solutions are $\vec{k}=0$ we have the identity (\ref{di2}).

\end{document}